\documentclass[letterpaper,english,aps,prb,floatfix,showpacs,twocolumn,amsfonts,amssymb,superscriptaddress,longbibliography]{revtex4-2}
\usepackage{balance}
\usepackage[T1]{fontenc}
\usepackage[applemac]{inputenc}
\usepackage[english]{babel}
\usepackage{graphics}
\usepackage{amssymb}
\usepackage{amsmath}
\usepackage{overpic}
\usepackage[toc]{appendix}

\usepackage{xcolor}
\usepackage{soul}
\usepackage{commath}
\usepackage{tikz}
\usepackage{physics}

\usepackage{hyperref}

\newcommand{\be}{\begin{equation}}
\newcommand{\ee}{\end{equation}}
\newcommand{\bspl}{\begin{split}}
\newcommand{\espl}{\end{split}}
\newcommand{\bea}{\begin{eqnarray}}
\newcommand{\eea}{\end{eqnarray}}

\newcommand{\bd}{\boldsymbol}

\def\a{\alpha}

\def\e{\varepsilon}
\def\d{\delta}
\def\g{\gamma}

\def\m{\mu}
\def\l{\lambda}

\def\t{\tau}

\def\o{\omega}

\def\s{\sigma}

\def\ct{\tilde c}

\def\D{\Delta}
\def\O{\Omega}
\def\L{\Lambda}

\def\ra{\rightarrow}

\def\pll{\parallel}

\def\pd{\partial}
\def\nb{\nabla}
\def\bdnb{\bd{\nb}}
\def\bk{{\bf k}}

\def\bq{{\bf q}}
\def\br{{\bf r}}

\def\bA{{\bf A}}

\def\bB{{\bf B}}
\def\bE{{\bf E}}

\def\bJ{{\bf J}}

\def\by{{\bf y}}

\def\bo{{\bf 0}}
\def\tc{\tilde{c}}
\def\tchi{\tilde{\chi}}

\def\tchi{{\tilde{\chi}}}
\def\hv{{\hat{\bd{v}}}}

\def\nn{\nonumber}
\def\lb{\label}
\def\pref#1{(\ref{#1})}

\newcount\bozza \bozza=0
\ifnum\bozza=1
\newdimen\shift \shift=-2truecm
\def\lb#1{%
{\label{#1}\rlap{\kern\shift{$\scriptstyle#1$}}}}
\else\def\lb#1{\label{#1}} \fi

\definecolor{darkred}{rgb}{0.55, 0.0, 0.0}
\definecolor{darkpowderblue}{rgb}{0.0, 0.2, 0.6}

\usetikzlibrary{shapes.arrows, fadings}
\tikzfading[name=fadeleft,
  left color=transparent!80, right color=transparent!0]

\begin{document}
\title{Charge density response in layered metals: retardation effects, generalized plasma waves and their spectroscopic signatures}
\author{F. Gabriele}
\email{francesco.gabriele@uniroma1.it}
\affiliation{Department of Physics and ISC-CNR, ``Sapienza'' University of Rome, P.le A. Moro 5, 00185 Rome, Italy}
\author{R. Senese}
\affiliation{Department of Physics and ISC-CNR, ``Sapienza'' University of Rome, P.le A. Moro 5, 00185 Rome, Italy}
\affiliation{\textnormal{Present address: }The Rudolf Peierls Centre for Theoretical Physics, Oxford University, Oxford OX1 3NP, UK}
\author{C. Castellani}
\affiliation{Department of Physics and ISC-CNR, ``Sapienza'' University of Rome, P.le A. Moro 5, 00185 Rome, Italy}
\author{L. Benfatto}
\email{lara.benfatto@roma1.infn.it}
\affiliation{Department of Physics and ISC-CNR, ``Sapienza'' University of Rome, P.le A. Moro 5, 00185 Rome, Italy}

\begin{abstract}
Transverse plasma polaritons and longitudinal plasmons describe the propagation of light-matter modes in an isotropic  metal. However, in a layered metal the anisotropy of the bare electromagnetic response mixes the longitudinal and transverse excitations, making the distinction between polariton and plasmon blurred at small wavevectors, where retardation effects of the electromagnetic interactions become quantitatively relevant. In the usual Kubo approach for the linear response, this effect appears as a mixing between the density and the transverse current fluctuations, that requires to revise the standard RPA approach for density correlations where only the instantaneous Coulomb potential is included. In this paper we derive the general expression for the density and current correlation functions at long wavelength in a layered metal, showing that below a crossover scale set by the anisotropy of the plasma frequencies retardation effects make the dispersion of the generalized plasma modes different from the standard RPA result. In addition, the mixed longitudinal and transverse nature of these excitations reflects in a double-peak structure for the density response, that can be eventually accessed by means of high-momentum resolution electron-energy-loss or X-rays spectroscopies. 
\end{abstract}
\date{\today}

\maketitle

\section{Introduction}

The propagation of electromagnetic (e.m.) waves in metals represents one of the main knobs to investigate the collective properties of the electronic system. On general grounds, in an isotropic metal transverse electromagnetic waves hybridize with the conduction-electron excitations giving rise to the so-called plasma polaritons, normal modes of the system propagating at a renormalized light velocity in the region of positive permittivity\cite{maier}. For zero momentum the frequency of plasma polaritons coincides with the frequency of longitudinal bulk plasmons, which are characterized by zero magnetic field and longitudinal electric field ($\bdnb \times \bE=\bo$), so that they satisfy Maxwell's equations under the condition of vanishing permittivity \cite{maier}. While plasma polaritons can be excited by an external e.m. radiation, longitudinal plasmons couple efficiently to density fluctuations and as such they are measured via electron-energy loss spectroscopies\cite{abajo_review} (EELS) or resonant inelastic X-Ray scattering\cite{vanderbrink_review} (RIXS), which access the charge-density response  of the electron gas \cite{pines,fetter}. Nowadays, the significant advances in the spectroscopic techniques using either confined light, as e.g. in. near-field optics \cite{schuller_review_nfo,basov-review-polaritons}, or integrating EELS with scanning transmission electron microscopy\cite{abajo_review,mitrano_pnas18,pichler_nature19}, made possible a detailed investigation of the energy-momentum dispersion of plasma modes at various length scales. A particular attention has been put to the wide category of layered metals, ranging from  van der Waals materials\cite{basov-review-polaritons,koppens-review-polaritons}  to layered high-$T_c$ cuprate superconductors\cite{keimer_review15}. 
%In the former case the interest is mainly motivated by the fact that in van der Waals materials the polariton wavelength can often be tuned with various methods, like e.g. electrical gating, making them a preferential platform for several applications based e.g. on the properties of surface plasma modes established at the interface with a dielectric.  

From the theoretical point of view, the behavior of metallic plasmons in a layered geometry is actually a very old problem, that has been studied since the late 70thies in connection with the physics of semiconducting superlattices \cite{grecu_prb73,fetter_ap74,dassarma_prb82,tselis_prb84}. The basic observation has to do with the fact that when conduction in the stacking-layers direction is poor the lack of screening of the interplane Coulomb interactions strongly modifies the plasmon dispersion with respect to the bulk isotropic case. For zero inter-layer momentum $q_z=0$, with $z$ being the stacking direction, one recovers the standard weakly dispersing plasmon as a function of the  momentum $q_\pll$ in the $xy$ plane, with a large (of order of the eV) value $\omega_{xy}$ at $q_\pll=0$. However, at finite $q_z$ the dispersion changes drastically, with a severe softening of the plasma energy towards $q_\pll=0$. In the limit of zero inter-layer hopping the plasmon energy goes to zero as $\sim q_\pll$ at finite $q_z=\pi/d$\cite{fetter_ap74,dassarma_prb82,tselis_prb84}, $d$ being the inter-layer distance, or it reaches a finite value of order of $\omega_z\ll\omega_{xy}$ when a finite hopping is allowed\cite{grecu_prb73}. With the discovery of high-temperature superconductivity in layered cuprates, that are well modelled by a stacking of weakly-coupled layers in the metallic phase, it has been also explored the  possibility that such "acoustic-like" plasmon branches can play a role in the superconducting phenomenon itself\cite{krezin_prb88,ishii_prb93,krezin_prb03}. However, the direct detection of the dispersive plasmon branches remained elusive for long time, mostly because in hole-doped cuprates RIXS signal is dominated by the spin excitations, that emerge strongly in proximity of the Mott-insulating antiferromagnetic phase. Only recently the existence of acoustic-like plasmon branches has been proven by RIXS, first in different families of electron-doped cuprates\cite{lee_rixs_nature18,liu_rixs_npjqm20} and more recently also in hole-doped cuprates\cite{zhou_prl20,huang_rixs_prb22}. In all these cases the plasmon dispersion follows qualitatively the prediction of a weakly-correlated layered electron model\cite{markiewicz_prb08,greco_prb16,phillips_prb22}, even though it does not capture the significant broadening of the plasmon observed at increasing in-plane momentum. Such an effect is even more pronounced in recent EELS measurements of the plasma dispersion at larger momenta\cite{abbamonte_scipost17,mitrano_pnas18,mitrano_prx19,abbamonte_cm23}, such that they have been interpreted as signatures of the so-called "strange metal" regime\cite{abbamonte_natcomm23}. At the same time, 
substantial work has been devoted in recent years to the possibility to drive non-linearly the soft inter-layer plasmon of cuprates at $\omega_z$ of few Thz with strong THz light pulses\cite{nori_review10,cavalleri_review,gabriele_natcomm21}. Indeed, from one side the gap opening below  the superconducting critical temperature $T_c$ makes it undamped\cite{uchida_prl92,homes_prl93,kim_physicac95,basov_prb94,vandermarel96}, in contrast to what happens in the metallic phase. From the other side, in the superconducting state the plasma modes appear also in the spectrum of the superconducting phase of the complex order parameter, allowing for its non-linear driving via optical probes\cite{nori_review10,cavalleri_review}. 

In the theoretical work aimed at describing the plasma modes measured by RIXS or EELS the usual approach \cite{grecu_prb73,dassarma_prb82,tselis_prb84,lee_rixs_nature18,liu_rixs_npjqm20,zhou_prl20,huang_rixs_prb22,phillips_prb22} consists in 
computing the density response of the anisotropic electron system by including at RPA level the effect of a  Coulomb-like interaction term, in analogy with the usual isotropic case\cite{vignale,fetter,pines}. This means that retardation effects, corresponding to the coupling of the charge density to the magnetic field induced via current fluctuations, have not been included\cite{fetter_ap74}. In the isotropic case this is actually not an approximation, but an exact result: indeed due to the complete decoupling between longitudinal and transverse degrees of freedom density fluctuations only induce longitudinal current fluctuations, remaining then decoupled from the magnetic field. On the other hand, in a layered system the anisotropy of the current response with respect to the in-plane and out-of-plane directions leads to an unavoidable coupling between charge and transverse current fluctuations, making magnetic-field effects in general not zero\cite{bulaevskii_prb94,bulaevskii_prb02}. The main consequence of the imperfect longitudinal/transverse decoupling is that in a layered metal there is an intrinsic mixture among plasmon and polaritons at generic wavevector, while an almost perfect decoupling is only reached at a momentum scale larger than a threshold $q_c\sim \sqrt{\omega_{xy}^2-\omega_z^2}/c$, set by the plasma-mode anisotropy\cite{gabriele_prr22,sellati_prb23}. Above this scale retardation effects are irrelevant and the standard approach including only Coulomb interactions between electrons is quantitatively correct. For typical values of $\omega_{xy}$ and $\omega_z$ the scale is $q_c\sim1-10$ $\mu$m$^{-1}$, so it is much smaller than the state-of-the art momenta accessible by RIXS and EELS. However, for the experiments with THz light mentioned above the regime $q<q_c$ is being probed, and indeed the role of magnetic-field effects has been discussed within the recent literature focusing on THz driving of the soft inter-layer plasmon below $T_c$
\cite{bulaevskii_prb94,bulaevskii_prb02,machida_prl99,machida_physc00,nori_review10,cavalleri_review,nori_natphys06,demler_prb20,demler_cm21}. It is worth noting that the theoretical investigation of plasma modes in the superconducting state is actually easier than in the metal, since plasmons appear in the response of the superconducting phase, whose dynamics has a relatively simple description at long wavelength\cite{gabriele_prr22,sellati_prb23}. In a recent publication\cite{gabriele_prr22} three of us took advantage of this peculiarity to derive an analytical expression for the generalized plasma modes in the SC state of layered superconductors via an effective-action formalism for the phase degrees of freedom, further extended to the bilayer case in Ref.\ \cite{sellati_prb23}. In this manuscript we aim at providing an analogous derivation for the layered metal, by employing again an effective-action formalism where both Coulomb and retardation effects are taken into account by integrating out the fluctuations of the internal e.m. degrees of freedom. This procedure is formally equivalent to an RPA approximation for both the density-density and current-current interactions, mediated respectively by the Coulomb potential and the transverse e.m. propagator. This approach allows us to derive in a rather compact and elegant way the density and current response to an external perturbation valid at any momenta, in terms of the bare susceptibility of the layered system. Such a formulation has the twofold advantage to allow us  for an analytical derivation of the generalized plasma waves for the uncorrelated layered metal, and to provide us with a general expression where short-range correlation effects can be included in the electronic susceptibilities. Indeed, as long as short-range interactions are included by preserving the gauge-invariance relations for the electronic response functions\cite{katsnelson_prl14,parcollet_prb14},  our scheme allows one to derive the plasma modes including retardation effects. 
 As an example we study specifically the density response, as accessed by EELS and RIXS experiments, and we show that the coupling among longitudinal and transverse degrees of freedom leads to a doubling of the peaks of the loss function at small momenta, that can eventually become accessible by improving momentum resolution in these probes.
 
The plan of the paper is the following. In Sec.\ \ref{retmax} an introductory analysis of the influence of retardation effects on the longitudinal propagation of plasmons is provided within the framework of Maxwell's Equations. As anticipated before, while in isotropic systems plasmons are never influenced by those effects, in layered systems the anisotropy of the bare electronic response leads to a non-trivial mixing between instantaneous (longitudinal) and retarded (transverse) fields. The same issue is then analyzed in detail within a microscopic many-body approach in Secs. \pref{iso} and \pref{ani}. In Sec.\ \ref{iso}, in order to give a pedagogical illustration of the formalism employed throughout the manuscript, we consider the linear response of an isotropic electron gas in the absence and in the presence of e.m. interactions, finding in both cases the results usually discussed in the literature: in the first case we find the Lindhard response functions, which are then renormalized at a standard RPA level when e.m. interactions are included. In Sec.\ \ref{ani} we address the linear-response theory of a layered system: we find that the mixing discussed in Sec.\ \ref{retmax} has the crucial consequence that the response functions deviate, at low momenta, from their standard-RPA counterparts. A remarkable consequence is that the correct density-density response function accounts for the propagation of two mixed longitudinal-transverse modes, which both appear in the spectrum of density fluctuations, that we discussed in detail in Sec.\ \ref{spec} for the case of a weakly-correlated layered metal. We then conclude the paper with a a general discussion about the results in Sec.\ \ref{conc}.

\section{Retardation effects and transverse/longitudinal mixing from Maxwell's equations}\lb{retmax}

To outline the physical mechanism behind the propagation of e.m. modes in a layered metal, that will be addressed in the rest of the paper by using a general many-body formalism, we start from the framework of classical Maxwell's Equations, connecting the density $\rho$ and currents  $\bJ$ fluctuations to the electric  $\bE$ and magnetic $\bB$ fields. The former is defined by the Gauss' law and Faraday's law as\cite{griffiths}
\bea
\lb{E}
\bdnb\cdot\bE(\br,t)=4\pi\rho(\br,t),\quad
\bdnb\times\bE(\br,t)=-\frac{1}{c}
\frac{\pd\bB}{\pd t}(\br,t),
\eea
where $c$ is the light velocity in vacuum, while the magnetic field is defined by the divergence-free condition along with the Ampere-Maxwell equation
\bea
\lb{B}
\bdnb\cdot\bB(\br,t)=0,
\quad
\bdnb\times\bB(\br,t)=
\frac{4\pi}{c}\bJ(\br,t)+
\frac{1}{c}
\frac{\pd \bE}{\pd t}(\br,t).\nn\\
\eea
The above equations can also be formulated in terms of scalar $\phi$ and vector $\bA$ potentials as
\be
\lb{fieldspot}
\bE(\br,t)=-\bdnb\phi(\br,t)-\frac{1}{c}
\frac{\pd\bA}{\pd t}(\br,t),\text{ }
\bB(\br,t)=\bdnb\times\bA(\br,t),
\ee
that make explicit the fact that while $\bB$ is always a transverse field, the electric field $\bE=\bE_L+\bE_T$ has in general both a longitudinal $\bdnb\times\bE_L=\bo$  and a transverse $\bdnb\cdot\bE_T=0$ component. As it is discussed in standard textbooks\cite{jackson}, the $\bE_T$ component, that according to Faraday's law \pref{E} is induced by a time-varying magnetic field, is the main responsible for the retardation effects, i.e. it mediates an interaction that takes a finite amount of time $\D t\equiv|\br-\br'|/c$ to propagate to point $(\br,t)$ from the source at $(\br',t-|\br-\br'|/c)$. This can be easily seen e.g. by using the 
Coulomb gauge $\bdnb\cdot\bA=0$, where $\bA\equiv\bA_T$ is purely transverse. In this gauge, that we will also use in the derivation below, the equations for the scalar and vector potentials decouple. The scalar potential satisfies the same Poisson equation $\nb^2\phi=-4\pi\rho$ of the electrostatic, so that the longitudinal electric-field component $\bE_L=-\nb\phi$ is not retarded. However, in the same gauge the vector potential $\bA_T$ satisfies an inhomogeneous d'Alembert equation with the transverse component of the current $\bJ_T$ as the only source, and whose solution is a retarded potential\cite{jackson}
\bea
\lb{ret}
\bA(\br,t)=
\frac{1}{c}\int d^3 r'
\frac{\bJ_T(\br', t-|\br-\br'|/c)}{|\br-\br'|}
\eea
One then sees that the $\bE_T$ component, expressed by Faraday's law as $\bE_T=-(\pd \bA_T/\pd t)/c$, accounts for retardation effects and vanishes as $c\ra\infty$: for this reason, such a feedback of the current fluctuations on the electric field is sometimes referred to as a "relativistic" effect. 

The previous discussion does not include yet the effect, specific of metals, of the current induced as local response to an electric field. As we shall clarify, such an induced response is responsible for the mixing between longitudinal and transverse e.m. modes in a layered metal. We consider first an isotropic conducting medium where $\bJ=\s\bE$, with $\s$ being a scalar conductivity, so that $\bJ$ and $\bE$ are parallel, and we switch for convenience to a Fourier-space notation. Let us assume that the external perturbation induces a finite charge fluctuation $\rho$, which in turn induces a longitudinal electric field with magnitude $|\bE_L|=(4\pi/|\bq|)\rho$, as given be Gauss's Law. Since 
$\bJ$ is parallel to $\bE$, $\bE_L$, as due to $\rho$, can only induce a longitudinal current $\bJ_L$. This implies that we cannot have any source for the magnetic field in Ampere-Maxwell's equation and, therefore, any finite transverse field $\bE_T$ from Faraday's Law. Finally, by approximating the conductivity with the Drude model at frequencies larger than the inverse electronic scattering rate as $\s\simeq -\o_p^2/(i 4\pi\o)$, $\o_p\equiv\sqrt{4\pi e^2 n/m}$ being the 3D plasma frequency, and using the continuity equation $\pd_t\rho=-\bdnb\cdot\bJ_L$, we end up with a closed equation for $\bE_L$:
\be
\lb{El}
\left(1-\frac{\o_p^2}{\o^2}\right)|\bE_L(\bq,\o)|=0.
\ee
The solutions of Eq.\ \pref{El} with $|\bE_L|\neq 0$ require $\o=\o_p$, that is the (dispersionless) expression for the longitudinal plasma mode in an isotropic conductor. In isotropic systems retardation effects do not affect longitudinal excitations, i.e. plasmons, since a longitudinal electric field never induces a transverse current as a source of a magnetic field. Conversely, a transverse current perturbation, induced in the metal in response to external transverse waves in the vacuum, does not induce any longitudinal response, making the polariton propagation independent from the plasmon.

In layered materials the situation is radically different. The electronic excitations in these systems can be modelled, in first approximation, with anisotropic effective masses for propagation in the planes or perpendicular to them, i.e. $m_{xy}\neq m_z$. This results in an anisotropy of the conductivity tensor given, in Cartesian coordinates, by $\hat{\s}=\begin{pmatrix} \s_{xy} & 0 \\ 0 & \s_z \end{pmatrix}$, with $\s_{xy/z}\simeq \o_{xy/z}^2/(i 4\pi\o)$, $\o_{xy/z}\equiv\sqrt{4\pi e^2 n/m_{xy/z}}$ being the plasma frequency along the $xy$ plane/$z$ axis, so that in general $\bJ$ and $\bE$ are no more parallel. For a perturbation with momentum $\bq$ forming a generic angle $\eta$ with  the $z$-axis we obtain, by simple rotation to the longitudinal/transverse basis, the general relation between the current and the electric field\cite{gabriele_22sif} as
\be
\lb{JEmatr}
\begin{pmatrix}
\bJ_L
\\
\bJ_T
\end{pmatrix}=
\begin{pmatrix}
\s_L & \s_{mix}
\\
\s_{mix} & \s_T
\end{pmatrix}
\begin{pmatrix}
\bE_L
\\
\bE_T
\end{pmatrix}
\ee
where $\s_{L/T}=\left(\s_{xy}q_{xy/z}^2+\s_zq_{z/xy}^2\right)/|\bq|^2$ is the longitudinal/transverse part of  the conductivity tensor and the non-diagonal element is defined as
\bea
\lb{smix}
\s_{mix}&=&
-\left(\s_{xy}-\s_z\right)
\frac{q_x q_z}{|\bq|^2}\nn\\
&=&
-\frac{\s_{xy}-\s_z}{2}
\sin(2\eta).
\eea
Therefore if we now introduce, as in the isotropic case, a charge-density perturbation that induces a longitudinal electric field, a transverse current $\bJ_T=\sigma_{mix}\bE_L$ is also produced. $\bJ_T$ acts as a source for the magnetic field in the Ampere-Maxwell's equation  $\bB=4\pi/(i c|\bq|)\bJ_T=4\pi/(i c|\bq|)\s_{mix}\bE_L$, where we neglected in first approximation the contribution of the displacement current. At the end, a {\em transverse}  electric field $\bE_T=(\o/c|\bq|)\bB$, as prescribed by Faraday's Law, appears in response to a {\em longitudinal} perturbation. The relative magnitude among the two components $|\bE_T|/|\bE_L|$ is approximately given by
\be
\lb{ratiolt}
\frac{|\bE_T|}{|\bE_L|}=
\frac{4\pi\o}{c^2|\bq|^2}|\s_{mix}|
\simeq
\frac{q_c^2}{|\bq|^2}
\frac{\sin(2\eta)}{2},
\ee
where we defined the momentum $q_c$ as
\be
\lb{cross}
q_c\equiv
\frac{\sqrt{\o_{xy}^2-\o_z^2}}
{c}.
\ee
A better estimate of the ratio among the longitudinal and transverse components within the Maxwell's formalism is presented in Appendix \ref{appa}, and an analogous one will be derived below within the Many-Body approach. Nonetheless, Eq.\ \pref{ratiolt} already gives an idea of the mechanism at play in layered systems. First of all,  Eq.\ \pref{ratiolt} is zero  for purely in-plane ($\eta=\pi/2$) or out-of-plane propagation ($\eta=0$), showing that no mixing occurs in these cases. 
 At a generic angle, according to the same equation, when $|\bq|\gg q_c$ one can neglect the induced transverse electric field, and thus retardation effects, so that the plasmon decouples from the polariton and one recovers the result of the isotropic case. In the Many-Body language, we expect this to be the regime where transverse current fluctuations induced by a density perturbation are negligible. Conversely, when $|\bq|\sim q_c$ one must account for retardation effects and longitudinal and transverse modes become intrinsically mixed. Eq.\ \pref{ratiolt} allows one to estimate the relevance of such effects depending on the probe under consideration, that sets the value of the momentum $\bq$. In systems like cuprates typical values of the plasma frequencies are $\o_{xy}\sim 1$ eV and $\o_z\sim 10^{-3}\o_{xy}$. Using $\hbar c\sim1.9\cdot 10^2$ eV nm one obtains that the largest value of the crossover momentum is $q_c\sim10^{-3}-10^{-2}$ nm$^{-1}$. At present the typical momentum resolution of RIXS does not exceed $\sim 0.1$ nm$^{-1}$, and it can be even larger for EELS, pushing then the measurement in a regime where retardation effects cannot be appreciated. Conversely, for light propagation the momentum is set by the frequency of the probe, being $q=\omega/c$. It then turns out that the maximum value of $|\bE_L|/|\bE_T|$ scales as $\sqrt{\omega_{xy}^2-\omega_z^2}/\omega\simeq \omega_{xy}/\omega$, where $\omega$ is the frequency of the e.m. radiation. One then understands why the mixing is crucial for $\omega$ of the order of few THz (1 THz$\simeq 4.1$ meV), as indeed discussed within the context of layered superconductors\cite{bulaevskii_prb94,bulaevskii_prb02,machida_prl99,machida_physc00,nori_review10,cavalleri_review,nori_natphys06,demler_prb20,demler_cm21}. In the next Sections the above results will be derived within a many-body approach to  linear-response theory, with the aim of providing a general structure of the density response in a layered metal that includes retardation effects when needed, and can be extended to the case of correlated metals, where also short-range interactions play a crucial role in determining the electronic response.

\section{Response Functions for the Isotropic System}\lb{iso}

\subsection{Path-integral approach to linear-response theory}

In order to provide a pedagogical illustration of the formalism, we consider the case of an isotropic free-electron gas, which is widely discussed in textbooks in the context of Many-Body Green's-function formalism \cite{nagaosa, fetter}. For the sake of simplicity, we put $\hbar=k_B=1$ in the following. We start from the imaginary-time action for the non-interacting electrons, which reads, in real and Fourier space respectively,\\
\bea
\lb{sfree}
S_0[\overline{\psi},\psi]&=&
\sum_{\s}
\int_0^{\frac{1}{T}} d\tau 
\int d\br
\overline{\psi}_\s(\br,\t)\nn\\
&\cdot&\left(\pd_\t-
\frac{\bdnb^2}{2m}-\mu
\right)\psi_\s(\br,\t)\nn\\
&=&-\sum_{k}\sum_{\s} \overline{\psi}_\s(k)
\mathcal{G}_0^{-1}(k)
\psi_{\s}(k),
\nn\\
\eea
where $\s$ is the spin index and $k$ is a shortcut for the momentum $\bk$ and the fermionic Matsubara frequency $\o_l=(2l+1)\pi T$. In the last row of Eq. \pref{sfree} we introduced the free-electron Matsubara Green's function $\mathcal{G}_0(k)=1/\left(i\o_l-\xi_{\bk}\right)$, where $\xi_\bk=|\bk|^2/(2m)-\mu$ is the free-electron energy dispersion with respect to the chemical potential $\mu$, $m$ being the effective mass of the electron. Since we are interested in computing the electromagnetic (e.m.) response we introduce the scalar and vector potentials $\phi$ and $\bA$ associated with the e.m. fields by means of the usual minimal-coupling substitution on both time and space derivatives as \cite{nagaosa, mandlshaw}%({\bf nota: $i\pd_t \rightarrow i\pd_t - q_e\phi(\br,t)$ and $-i\bdnb \rightarrow-i\bdnb-(q_e/c)\bA(\br,t)$, ossia $i\pd_\mu \ra i\pd_\mu - q_e A_\mu$, infine poni $q_e=-e$})
\be
\lb{mincoup}
i\pd_\mu \ra i\pd_\mu - 
\frac{e}{c} A_\mu,
\ee
where $e>0$ is the absolute value of the electron charge, $\pd_\mu=(\pd_t,\bdnb)$ is the 4-gradient operator and we introduced 
$A^{\mu}=(c\phi,\bA)$ and $A_{\mu}=(-c\phi,\bA)$ as the contravariant and covariant 4-potentials, respectively. Eq. \pref{mincoup} ensures that the total action exhibits an invariance under simultaneous gauge transformations for the fermionic and the e.m. potentials, i.e.
\be
\lb{gautran}
\begin{cases}
& \psi(\br,t) \ra\psi(\br,t)
\exp\left(-\frac{i e}{c}
\l(\br,t)\right),
\\
\\
& A_\mu(\br,t)\ra A_\mu(\br,t)+\pd_\mu\l(\br,t),
\end{cases}
\ee
%
%({\bf nota: in spazio reale $\phi\ra\phi-(1/c)\pd_t\l$ e $\bA\ra\bA+\bdnb\l$, oppure, moltiplicando la seconda per $1/c$, $\bA/c\ra\bA/c+(1/c)\bdnb\l$. Dunque, con i 4-vettori, ho $A_\mu\ra A_\mu-(1/c)\pd_\mu$, da cui poi l'espressione 4-vettoriale in spazio di Fourier}).
where $\l$ is an arbitrary function. In the imaginary-time formalism where $it\rightarrow \tau$ one equivalently replaces $\pd_\tau \ra \pd_\tau-e\phi$. Since the charge density and current are defined, as usual, as functional derivatives of the action with respect to the e.m. potentials, one can express the induced 4-current $J^\mu=(\rho,\bJ)$ ($\rho$ and $\bJ$ being, respectively, the induced density and current) to an external source field in linear-response theory as
\be
\lb{defjmu}
J^\mu(q)=-\frac{e^2}{c} K^{\mu\nu}(q)A_{\nu}(q),
\ee
where $q=(\bq,i\O_n)$ is a compact shortcut notation for the momentum $\bq$ and the bosonic Matsubara frequency $i\O_n=i2n\pi T$ and the response function $K^{\mu\nu}$ can be readily obtained as
\be
\lb{derk}
K^{\mu\nu}(q)=- 
\frac{c^2}{e^2}
\frac{\d^2 \ln Z [A] }{\d A_\mu(q) \d A_\nu(-q)}\vert_{A_\mu, A_\nu=0},
%\frac{\pd^2 S_{eff}[A_\mu]}{\pd A_\mu(q) \pd A_\nu(-q)}\vert_{A_\mu, A_\nu=0},
\ee
where the partition function reads $Z[A]=\int\mathcal{D}[\psi,\overline{\psi}] e^{-S[\psi,\overline{\psi},A_\mu]}$, $S$ being the imaginary-time action describing the quantum dynamics of the fermions in the presence of the e.m. fields. In a charged system the e.m. fields induce charge and density fluctuations within the medium that should be included in the density response, that is the main focus of the present work. In the usual perturbative approach one accounts for this effect by adding an interaction term in the electronic Hamiltonian accounting for density-density or current-current interactions. Here we will follow a different but completely equivalent approach, by making an explicit distinction between the internal statistical potentials, which, from now on, will be denoted by $A_\mu=(-c\phi,\bA)$, and the auxiliary external "source" fields, denoted by $A_\mu^{ext}=(-c\phi^{ext},\bA^{ext})$. In this case one can define the response to the external perturbation as
\be
\lb{derkext}
K_{ext}^{\mu\nu}(q)=-
\frac{c^2}{e^2}
\frac{\d^2 \ln Z [A^{ext}] }
{\d A_\mu^{ext}(q) \d A_\nu^{ext}(-q)}
\vert_{A_\mu^{ext}, A_\nu^{ext}=0},
%\frac{\pd^2 S_{eff}[A_\mu]}{\pd A_\mu(q) \pd A_\nu(-q)}\vert_{A_\mu, A_\nu=0},
\ee
where $Z[A^{ext}]=\int\mathcal{D}[\psi,\overline{\psi},A] e^{- S[\psi,\overline{\psi},A,A^{ext}]}$.
The integration over the internal e.m. degrees of freedom will account for the e.m. interaction among the electrons, and the 4-current \pref{defjmu} will be the response to the external perturbation. 

In order to highlight the role of the e.m. interactions and to make a direct analogy with known results, let us first neglect the effect of the internal e.m. fields and let us just compute the response to the external sources. Once introduced $A^{ext}$ by means of the prescription \pref{mincoup} we get the action
\be
\lb{stot}
S[\overline{\psi},\psi,A^{ext}]=S_0[\overline{\psi},\psi]
+S_{el+e.m.}
[\overline{\psi},\psi,A^{ext}],
\ee
where the coupling between the electrons and the auxiliary e.m. fields is encoded into $S_{el+e.m.}$, which reads
\bea
\lb{selem}
&&S_{el+e.m.}
[\overline{\psi},\psi,A^{ext}]=\nn\\
&=&\frac{e}{c}
\sum_\s\sqrt{\frac{T}{V}}
\sum_{k,k'}
\overline{\psi}_\s(k)
\psi_\s(k')
s^{\mu}(\bk,\bk')A_\mu^{ext}(k-k')
\nn\\
&+&\frac{e^2}{2 mc^2}
\sum_\s\frac{T}{V}
\sum_{k,k',q}
\overline{\psi}_\s(k)\psi_\s(k')
{\bA^{ext}(k-k'+q)\cdot\bA^{ext}(-q)},\nn\\
\eea
where $s^\mu(\bk,\bk')=\left(1,\left(\bk+\bk'\right)/(2m)\right)$ is the density-current vertex. Since Eq. \pref{stot} is quadratic in the fermionic fields, they can be integrated out exactly, leading to an effective action $S_{eff}[A^{ext}]$ which includes all powers of $A^{ext}$. However, in order to compute the response function $K^{\mu\nu}$ through Eq.\ \pref{derkext} it is sufficient to retain terms quadratic in the external fields, i.e. to define an effective Gaussian action $S_{G}$, such that
\be
\lb{sgauss}
S_{G}[A^{ext}]
=\frac{e^2}{2 c^2}\sum_q
A_\mu^{ext}(q) K^{\mu\nu}(q) A_\nu^{ext}(-q),
\ee
so that $Z=e^{-S_G}$ and $K^{\mu\nu}(q)\equiv 
\frac{c^2}{e^2}
\frac{\d^2 S_{G}[A^{ext}]}{\d A_\mu^{ext}(q)
\d A_\nu^{ext}(-q)}$, as a direct consequence of Eq.\ \pref{derkext}. It is worth noting that the response functions $K^{\mu\nu}$ cannot be independent on each other. Indeed, Eq. \pref{sgauss}, which depends on $A_\mu^{ext}$ only, must be still invariant with respect to the second transformation of Eq. \pref{gautran}, which reads, in Fourier space,
\be
\lb{gautranA}
A_\mu^{ext}(q)\ra A_\mu^{ext}(q)+i q_\mu \l(q),
\ee
where the 4-momentum is defined as $q_\mu=(-i\O_n,\bq)$. The gauge invariance requires that any additional term introduced into the action \pref{sgauss} by Eq.\ \pref{gautranA}, i.e. those proportional to $i q_\mu K^{\mu\nu} A_\nu^{ext}$, $i A_\mu^{ext} K^{\mu\nu} q_\nu$ and $q_\mu K^{\mu\nu} q_\nu$, must vanish. This is guaranteed only if the linear-response functions obey the following gauge-invariance conditions:
\be
\lb{gi}
q_\mu K^{\mu\nu}(q)=0,\quad\quad K^{\mu\nu}(q)q_\nu=0.
\ee
In the following we will check that both in the isotropic and anisotropic case such conditions are satisfied.

Let us then recall briefly the result obtained without the contribution of the internal e.m. fields, which is equivalent to the standard "bare" response of the non-interacting electron gas. It is straightforward to show that in this case the action \pref{sgauss} has the form
\bea
\lb{sgaussbare}
S^0_{G}[A^{ext}]
&=&\frac{e^2}{2 c^2}\sum_q
A_\mu^{ext}(q) \chi_0^{\mu\nu}(q) A_\nu^{ext}(-q)
\eea
where the bare susceptibilities $\chi_0^{\mu\nu}$ are the standard imaginary-time linear-response functions of the free-electron gas\cite{fetter}, i.e.
\be
\lb{bare}
\chi^{\mu\nu}_0(q)=
\frac{n}{m}
\d^{\mu\nu}\left(1-\d^{\mu0}\right)+\tchi_0^{\mu\nu}(q).
\ee
They are given by the sum of a diamagnetic-like term, i.e. the first one, and a paramagnetic-like one $\tchi_0^{\mu\nu}$, which is given by\cite{fetter}
\bea
\lb{para}
\tchi_0^{\mu\nu}(q)&=&\frac{2T}{V}\sum_{k}
\g^{\mu}\left(\bk,\bq\right)
\g^{\nu}\left(\bk,\bq\right)
\mathcal{G}_0(k+q)\mathcal{G}_0(k)=\nn\\
&=&\frac{2}{V}\sum_\bk 
\g^{\mu}\left(\bk,\bq\right)
\g^{\nu}\left(\bk,\bq\right)
\frac{f(\xi_\bk)-f(\xi_{\bk+\bq})}
{i\O_n-\left(\xi_{\bk+\bq}-\xi_{\bk}\right)}\nn\\
\eea
where the overall 2 factor accounts for the spin degeneracy and $f(\xi)=1/\left(\exp\left(\xi/T\right)+1\right)$ is the Fermi distribution. In Eq. \pref{bare}, $n=2(T/V)\sum_{i\o_l,\bk}\mathcal{G}_0(i\o_l,\bk)
e^{-i\o_l 0^-}=2/V\sum_\bk f(\xi_\bk)$ is the density of electrons (with spin degeneracy included). Also, in Eq. \pref{para}, the density-current vertex $\g^\mu$ is now defined as $\g^\mu(\bk,\bq)=\left(1,\left(\bk+\bq/2\right)/m\right)$. 

It is easy to prove that the bare response functions indeed fulfill the gauge-invariance conditions prescribed by Eq. \pref{gi}. To this aim, we notice that the isotropic bare density-current function is always longitudinal, i.e. parallel to $\bq$, since\cite{fetter}
\be
\lb{dciso}
\chi_0^{0i}(q)=
\frac{i\O_n q_i}{|\bq|^2}\chi_0^{00}(q)
\ee
and the isotropic current-current function always allows for the following longitudinal-transverse decomposition\cite{fetter}:\\
\be
\lb{cciso}
\chi_0^{ij}(q)=\frac{(i\O_n)^2}{|\bq|^2}
\chi_0^{00}(q)
\left(\hat{P}_L(\bq)\right)^{ij}+\chi_0^T(q)\left(\hat{P}_T(\bq)\right)^{ij}.
\ee
In Eq.\ \pref{cciso} $\chi^T_0=(1/2)\left(\hat{P}_T\right)_{ij}\chi_0^{ji}$ is the transverse part of $\chi_0^{ij}$, and $\hat{P}_L$ and $\hat{P}_T$ are, respectively, the longitudinal and transverse projection operators, defined as
\be
\lb{proj}
\left(\hat{P}_L(\bq)\right)_{ij}=\frac{q_i q_j}{|\bq|^2},\quad
\left(\hat{P}_T(\bq)\right)_{ij}=\d_{ij}-\frac{q_i q_j}{|\bq|^2}.
\ee
Given the two identities \pref{dciso} and \pref{cciso}, it is trivial to prove that $\chi_0^{\mu\nu}$ indeed satisfies the gauge-invariance conditions given by Eq. \pref{gi}, i.e.
\be
\lb{gibare}
q_\mu\chi_0^{\mu\nu}(q)=0 \quad\quad \chi_0^{\mu\nu}(q)q_\nu=0.
\ee
For instance, by virtue of Eqs.\ \pref{dciso} and \pref{cciso}, we have that $-i\Omega_m\chi_0^{00}+\chi_0^{0i}q_i=-i\Omega_m\chi_0^{00}+i\Omega_m \chi_0^{00}q_i q_i/|\bq|^2=0$ and $-i\Omega_m\chi_0^{0i}+\chi_0^{ij}q_j=-i\Omega_m\chi_0^{0i}+(i\Omega_m/|\bq|)^2 \chi_0^{00}q_i =0$, which are, respectively, the time-like ($\nu=0$) and the space-like ($\nu=i$) components of Eq.\ \pref{gibare}.

\subsection{Linear-response theory in the presence of electromagnetic interaction: isotropic systems}

Let us now show how the bare electronic response is dressed by the integration of the internal e.m. fields. To this aim, we couple the electrons to both internal and external fields by means of the minimal-coupling substitution \pref{mincoup}, and we add the e.m. action of the internal fields, which reads, in real and Fourier space respectively\cite{nagaosa, mandlshaw},
\bea
\lb{sem}
S_{e.m.}[A]
&=&\int d\t d\br 
\left[\frac{ (\bdnb\times\bA)^2 }{8\pi}
-\frac{\e}{8\pi} \left(\frac{i\pd_\t \bA}{c}+\bdnb \phi\right)^2\right] \nn\\
&=&\frac{\e}{8 \pi}
\sum_q \left[ -|\bq|^2 |\phi(q)|^2 \right. \nn\\
&+&\left(\O^2_m + \frac{c^2}{\e}|\bq|^2
\right)\frac{|\bA_T(q)|^2}{c^2}
+\O_m^2|\bA_L(q)|^2\nn\\
&+&\left.i\O_n\bq\cdot \left( \phi(q)\frac{\bA(-q)}{c}+\phi(-q)\frac{\bA(q)}{c}\right)
\right].
\eea
Eq. \pref{sem} is the transcription in imaginary time of the usual Lagrangian density $(-\e|\bE|^2+|\bB|^2)/(8\pi)$, where $\bB=\bdnb\times\bA$ is the magnetic field and the electric field $\bE$ reads $\bE=-(i/c)\pd_\tau\bA-\bdnb\phi$, with $\e$ being 
a background dielectric constant which accounts for ionic screening. It is worth noting that in order to have a definition of $\bE$ analogous to the one valid for real time one should assume that $\phi$ is purely imaginary, i.e. one should replace $\phi\ra i\phi$. In this case, by defining the imaginary-time electric field as $\bE\equiv-\frac{1}{c}\partial_\t\bA-\bdnb\phi$ the action for the free e.m. fields would read $\frac{\e|\bE|^2+|\bB|^2}{8\pi}$. Such a rescaling of the scalar potential would also make the quadratic term in the scalar potential arising from $\left(\bdnb\phi\right)^2$ positive defined, as required to perform the Gaussian integration. To make notation more compact we will not explicitly rescale the potential in what follows, but we will implicitly assume that a formal definition of the Gaussian integration in the imaginary-time formalism requires such a regularization. 
Finally, in Eq.\ \pref{sem} we introduced explicitly the longitudinal-transverse decomposition $\bA=\bA_L+\bA_T$ for the vector potential, where $\bA_L=\hat{\bq}\left(\hat{\bq}\cdot\bA\right)$ is the longitudinal part, such that $\bq\times\bA_L=\bo$, and $\bA_T=\bA-\bA_L=(\hat{\bq}\times\bA)\times\hat{\bq}$ is the transverse part obeying $\bq\cdot\bA_T=0$. This allows one to clearly identify the bare propagators for the internal gauge fields. Indeed, the coefficient of the quadratic term in $\phi$ can be recast as $-e^2/\left( 2 V_C(\bq) \right)$, where
\be
V_C(\bq)=\frac{4\pi e^2}{\e|\bq|^2}
\ee
is the Coulomb pontential, while $-\left(e/c\right)^2/(2 D_T)$ is the coefficient of the quadratic term in the transverse gauge field $\bA_T$, with
\be
\lb{phprop}
D_T(q)=
\frac{4\pi e^2/\e}
{\left(i\O_n\right)^2-\tc^2|\bq|^2},
\ee
being the transverse propagator. The poles of Eq. \pref{phprop} yield, after analytic continuation $i\O_n\ra\o+i 0^+$, the light dispersion $\o=\tc|\bq|$, where $\tc=c/\sqrt{\e}$ is the renormalized light velocity. Moreover, the last line of Eq.\ \pref{sem} shows that the scalar potential only couples to the longitudinal component of $\bA$, so that for isotropic systems the Coulomb gauge $\nb\cdot \bA=0$, i.e. $\bA_L=0$, allows one to completely decouple the scalar and the vector potentials. Since in this gauge the longitudinal part of the electric field $\bE_L=-\nabla \phi$  is controlled by the scalar potential only, one also achieves for the isotropic system a complete decoupling among  transverse and longitudinal degrees of freedom. However, as we shall see in the next section, in the anisotropic case even in the Coulomb gauge such a decoupling is not allowed. It is then more convenient for what follows to rewrite in Fourier space the $(\nb\times\bA)^2$ term as $|\bq\times A(\bq)|^2=|\bq|^2 |A(\bq)|^2-|\bq\cdot A(\bq)|^2$, so that  Eq.\ \pref{sem} reads
\begin{widetext}
\bea
\lb{sem2}
S_{e.m.}[A]&=&\frac{e^2}{2}
\sum_q \left[ -\frac{|\phi(q)|^2}{V_C(\bq)}
-\frac{|\bA(q)|^2/c^2}{D_T(q)}
-\frac{1}{4\pi e^2}|\bq\cdot\bA(q)|^2 +
\frac{\e }{4\pi e^2}i\O_n\bq\cdot
\left(\phi(q)\frac{\bA(-q)}{c}+
\phi(-q)\frac{\bA(q)}{c}\right)
\right].\nn\\
\eea
Once introduced the coupling of $A_\mu$ with the fermionic fields {according to Eq.}\ \pref{mincoup} we obtain the total action $S[\overline{\psi},\psi,A+A^{ext}]+
S_{e.m.}[A]$, where $S$ is given by Eq. \pref{stot}, apart from the fact that it now depends on the total field $A_\mu(q)+A_\mu^{ext}(q)$. Then, in full analogy with the free-electron case, we can integrate out the fermionic fields, which still appear at quadratic order. The result of the integration is twofold: from one side we recover the effect of matter on the bare e.m. response, and from the other we describe the perturbation due to the external source fields $A_\mu^{ext}$. The Gaussian action for both the internal and the external e.m. fields is explicitly given by
\begin{subequations}
\lb{sgaussint}
\begin{align}
S_G^{iso}[A, A^{ext}] &
=\frac{e^2}{2 c^2}\sum_q
\left( A_\mu(q)+A_\mu^{ext}(q) \right)
\chi_0^{\mu\nu}(q) 
\left( A_\nu(-q)+A_\nu^{ext}(-q) \right)
+S_{e.m.}[A]
\nn\\
\lb{sga}
&=\frac{e^2}{2}\sum_q\left[
\chi_0^{00}(q)|\phi^{ext}(q)|^2+
\chi_0^{ij}(q)
\frac{A_i^{ext}(q)}{c}
\frac{A_j^{ext}(-q)}{c}
-\chi_0^{0i}(q)\phi^{ext}(q)
\frac{A_i^{ext}(-q)}{c}
+c.c.
\right. \\
\lb{sgb}
&+
\chi_0^{00}(q)\phi^{ext}(q)\phi(-q)
+\chi_0^{ij}(q)
\frac{A_i^{ext}(q)}{c}
\frac{A_j(-q)}{c}
-\phi^{ext}(q)\chi_0^{0i}(q)
\frac{A_i(-q)}{c}
-\phi(q)\chi_0^{0i}(q)
\frac{A_i^{ext}(-q)}{c}
+c.c.\\
\lb{sgc}
&+\left(\chi_0^{00}(q)-\frac{1}{V_C(\bq)}\right)|\phi(q)|^2
+
\left(\chi_0^{ij}(q)-
\frac{1}{D_T(q)}\d_{ij}\right)
\frac{A_i(q)}{c}\frac{A_j(-q)}{c}
-\phi(q)\chi_0^{0i}(q)\frac{A_i(-q)}{c}+c.c.\\
\lb{sgd}
&+\left.
\frac{\e }{4\pi e^2}i\O_n\bq\cdot
\left(\phi(q)\frac{\bA(-q)}{c}+
\phi(-q)\frac{\bA(q)}{c}\right)
-\frac{1}{4\pi e^2}
\left(\bq\cdot\bA(q)\right)^2
\right].
\end{align}
\end{subequations}
\end{widetext}

As for the free-electron case, the invariance of the action \pref{sgaussint} under local-gauge transformations of both $A_\mu$ and $A_\mu^{ext}$ is ensured by Eq. \pref{gibare} for the bare response functions $\chi_0^{\mu\nu}$. As a last step  one integrates out the internal potential $A_\mu$ in Eq. \pref{sgaussint}, that is equivalent to compute the response functions at a RPA level in the usual diagrammatic approach to fermionic models\cite{fetter}. The inclusion in Eq.\ \pref{sgaussint} of higher-order terms in $A_\mu$ would yield, once integrated out, beyond-RPA corrections to the linear-response functions: these will not be addressed in the present work. Before solving the integral, one must fix the gauge for the internal potentials, in order to get rid of the divergence due to the redundancy of $A_\mu$: indeed, there are infinitely many potentials $A_\mu+i q_\mu \l$ accounting for the same physical configuration of the e.m. fields $\bE$ and $\bB$, and such an arbitrariness leads to a divergence of the functional integral. A proper gauge-fixing procedure prevents the Gaussian integral from being singular. For the isotropic case, the Coulomb gauge $\bdnb\cdot\bA=0$, i.e. $q_i A_i=0$ in Fourier space, makes the computations rather straightforward. Indeed, since the bare isotropic density-current function \pref{dciso} is always longitudinal, i.e. $\chi_0^{0i}\propto q_i$, it follows that all the terms proportional to $\chi_0^{0i}A_i\propto q_i A_i$ vanish in the Coulomb gauge. It then follows that the last two terms of Eq.\ \pref{sgb}, the last term of Eq.\ \pref{sgc} and the whole Eq.\ \pref{sgd} cancel out, leading to a complete decoupling among the scalar and the vector potentials, that can be integrated out separately to obtain the full response functions
$\tilde \chi_{iso}^{\mu\nu}$ of the isotropic case. Integrating out $\phi$ is then equivalent to the standard RPA dressing of the bare bubbles $\chi_0^{00}$, $\chi_0^{0i}$ and $\chi_0^{ij}$ with respect to the Coulomb potential:
\be
\lb{stadrpadd}
\tchi_{iso}^{00}(q)\equiv
\chi_{RPA}^{00}(q)=
\frac{\chi_0^{00}(q)}{1-V_C(\bq)\chi_0^{00}(q)},
\ee
\be
\lb{stadrpadc}
\tchi_{iso}^{0i}(q)\equiv
\chi_{RPA}^{0i}(q)=\frac{\chi_0^{0i}(q)}
{1-V_C(\bq)\chi_0^{00}(q)},
\ee
\be
\lb{stadrpacc}
\chi_{RPA}^{ij}(q)=\chi_0^{ij}(q)
+V_C(\bq)
\frac{\chi_0^{i0}(q)\chi_0^{0j}(q)}
{1-V_C(\bq)\chi_0^{00}(q)}.
\ee
First of all, we notice that, thanks to the structure encoded into Eqs.\ \pref{stadrpadd}-\pref{stadrpacc}, and to the relations \pref{gibare}, also the standard-RPA response functions obey the gauge-invariance conditions prescribed by Eq. \pref{gi}, i.e. $q_\mu\chi_{RPA}^{\mu\nu}=0$ and $\chi_{RPA}^{\mu\nu}q_\nu=0$. This is a direct consequence of the fact that, in the current-current function given by Eq. \pref{stadrpacc}, the standard-RPA correction has a purely longitudinal structure, i.e. it only renormalizes the longitudinal part $\chi_0^L$ of the bare current function, which already satisfies the condition \pref{gi}. This is in agreement with the observation done before that for the isotropic system the longitudinal degrees of freedom in the Coulomb gauge are fully described by the scalar potential. The most complete current-current function $\tchi_{iso}^{ij}$ is, in fact, obtained after integration of $\bA$ as well, that only dresses the transverse sector $\chi_0^T$ of $\chi_0^{ij}$ with respect to the transverse propagator $D_T$, due to the absence of coupling term between $\bA$ and $\phi^{ext}$. Once both integrations are carried out, one finds that the full current function $\tchi_{iso}^{ij}$ is given by
\be
\lb{chicciso}
\tchi_{iso}^{ij}(q)=\tchi_{iso}^L(q)
\left(\hat{P}_L(\bq)\right)^{ij}+
\tchi_{iso}^T(q)
\left(\hat{P}_T(\bq)\right)^{ij},
\ee
where  $\tchi_{iso}^L=\chi_{RPA}^L=\left(i\O_n/|\bq|\right)^2\tchi_{iso}^{00}$ equals the longitudinal part of Eq. \pref{stadrpacc}, while $\tchi_{iso}^T$ is given by
\be
\lb{isotcc}
\tchi_{iso}^T(q)=
\frac{\chi_0^T(q)}
{1 - D_T(q)\chi_0^T(q)}. 
\ee

Having computed the electronic response in the presence of internal e.m. fields, we now briefly recall the standard outcomes for the collective modes of an isotropic system, in which Eq.\ \pref{defjmu} reduces to the following two independent equations for the density $\rho$ and the transverse current $\bJ_T$:
\be
\lb{defjmuiso}
\rho(q)=\tchi_{iso}^{00}(q)\phi^{ext}(q),
\quad
\bJ_T(q)=\tchi_{iso}^T(q)
\bA_T^{ext}(q).
\ee
We did not mention the equation for the longitudinal current $\bJ_L=\tchi_{iso}^L\bA_L^{ext}$, since it carries the same information of the first of Eq. \pref{defjmuiso}, $\tchi_{iso}^L$ being proportional to $\tchi_{iso}^{00}$ and $\bJ_L$ being related to $\rho$ through the continuity equation. Within this context, the longitudinal plasmon and the transverse plasma polariton appear as poles of the density-density response $\tilde\chi^{00}_{iso}$ and of the transverse current response $\tchi_{iso}^T$ respectively. At long wavelength we can derive an analytical expression for both modes by using the approximated behavior\cite{fetter} of the Lindhard functions in the long-wavelength dynamical limit $v_F|\bq|\ll\o$ ($v_F$ being the Fermi velocity), such that
\bea
\lb{approx0}
\chi_0^{00}(q)&\simeq&
\frac{n|\bq|^2}{m\o^2}
\left(1+
\frac{3}{5}\frac{v_F^2|\bq|^2}{\o^2}
\right),\\
\lb{approxt}
\chi_{iso}^T(q) 
&\simeq& \frac{n}{m}.
\eea
As a consequence for the longitudinal mode the pole of Eq.\ \pref{stadrpadd} gives
\be
\lb{isodisp}
1-V_C(\bq)\chi_{00}(q)=0 \Longrightarrow 
\o_L(\bq)=\o_p\sqrt{1+v_p^2|\bq|^2},
\ee
with $v_p^2=3 v_F^2/(5\omega_p^2)$ setting the scale of the plasmon dispersion, where $\o_p$ is the 3D isotropic plasma frequency defined as
\be
\lb{pf}
\o_p\equiv\sqrt{\frac{4\pi e^2 n}{\e m}}.
\ee
On the other hand the plasma polariton is defined by the pole of Eq.\ \pref{isotcc}, and it is given by
\be
\lb{polar}
1 - D_T(q)\chi_0^T(q)
\Longrightarrow 
\o_T(\bq)=\sqrt{\o_p^2+\tc^2|\bq|^2}.
\ee
Notice that terms of order $v_F^2|\bq|^2$ have been neglected in the dispersion of the polariton \pref{polar} since as usual $v_F<<\tc$, being $\tc\sim 10^8$ ms$^{-1}$, far bigger than the typical Fermi velocity $v_F\sim 10^6$ ms$^{-1}$ of an isotropic metal.

A better insight onto the role of the plasmon for the density response is obtained by deriving the general expression for its spectral function $S(q)\equiv-\Im\tilde\chi_{iso}^{00}(q)$ as
\bea
\lb{specdens}
S(q)=-\frac{{\chi_0^{00}}''(q)}
{\left(1-V_C(\bq){\chi_0^{00}}'(q)
\right)^2
+\left( V_C(\bq)
{\chi_0^{00}}''(q) \right)^2
},\nn\\
\eea
where the single and the double apostrophes denote, respectively, the real and the imaginary parts of the bare density bubble, and the overall minus sign is due to the fact that we are considering retarded response functions\cite{fetter}. In the simplified case in which short-range interactions are negligible the long-wavelength dynamical bare density bubble has a vanishing imaginary part, i.e. ${\chi_0^{00}}''\ra 0^-$, due to the absence of particle-hole excitation at $v_F|\bq|\ll\o$ in a free-electron gas\cite{fetter,vignale}, while the real part can be once again expanded as ${\chi_0^{00}}'\simeq n|\bq|^2/(m\o^2)$. One then finds that within such a limit Eq. \pref{specdens} displays a delta-like peak centered around $\o_L$, i.e.
\be
\lb{plasmawidth}
S(q)
\simeq
\pi I_L(\bq)
\d \left(\o-\o_L(\bq)\right),
\ee
where the overall spectral weight is given by
\be
\lb{Il}
I_L(\bq)=
\frac{\o_L(\bq)}{2 V_C(\bq)}.
\ee
One then sees that the peak at the plasmon in the density response has zero spectral weight as $\bq\ra \bo$, as expected by charge conservation\cite{fetter}.

\section{Response functions for a Layered System}\lb{ani}

So far, we considered the case of an isotropic system: in this sense, we were allowed to consider the effective mass $m$ a scalar quantity. Here we will be interested instead in describing \textit{layered} materials, which are made of 2D weakly {coupled} conducting planes. Such a weak interaction between the layers strongly suppresses the out-of-plane transport: one can account for such an \textit{anisotropy} within an approximate free-electron continuum model with an effective mass $m_i$ depending on the direction $i$, with $m_i=m_{xy}$ for $i=x,y$, and $m_i=m_z$ for $i=z$, being $m_{xy}<m_z$. Substituting the isotropic mass $m$ with the index-dependent one $m_i$ yields the \textit{anisotropic} bare Lindhard bubbles, which we denote here by $\Pi_0^{\mu\nu}$. Their form can be easily derived from the known expansion of the isotropic case\cite{fetter} by mapping the anisotropic electron gas with effective masses $m_{xy}$ and $m_z$ into a fictitious isotropic one with effective mass $m^*\equiv\left(m_{xy}^2m_z\right)^{1/3}$. Such a procedure, which is shown in detail in Appendix \ref{appb}, leads to the following identities:
\be
\lb{poo}
\Pi_0^{00}(\bq,\o)=
\chi_{0*}^{00}(\tilde{\bq},\o)
\ee
\bea
\lb{pio}
\Pi_0^{0i}(\bq,\o)&=&
\sqrt{\frac{m^*}{m_i}}
\chi_{0*}^{0i}(\tilde{\bq},\o)\nn\\
&=&
\sqrt{\frac{m^*}{m_i}}
\frac{ i\O_n \tilde{q}_i }
{|\tilde{\bq}|^2}
\chi_{0*}^{00}(\tilde{\bq},\o)
\eea
\bea
\lb{pij}
\Pi_0^{ij}(\bq,\o)&=&
\sqrt{\frac{m^*}{m_i}}
\sqrt{\frac{m^*}{m_j}}
\chi_{0*}^{ij}(\tilde{\bq},\o)
\eea
The momentum $\tilde{\bq}$ is rescaled such that its components satisfy $\tilde{q}_i=\sqrt{\frac{m^*}{m_i}} q_i$, and $\chi_{0*}^{\mu\nu}$ denotes the generic response function of the isotropic free-electron gas with effective mass $m^*$. The second row of Eq.\ \pref{pio} has been rewritten, for the sake of the following discussion, by taking advantage of the expression of the isotropic density-current function provided by Eq.\ \pref{dciso}.

The rescaling encoded in Eq.s \pref{pio}-\pref{pij} does not affect the gauge-invariance condition of the non-interacting electron system, so that the $\Pi_0^{\mu\nu}$ function still satisfy Eq.\ \pref{gi}:
\be
\lb{piogi}
q_\mu \Pi_0^{\mu\nu}(q)=0,\quad\quad
\Pi_0^{\mu\nu}(q) q_\nu=0. 
\ee
On the other hand, from Eq.\ \pref{pio} it follows that, in contrast to the isotropic case where $\chi_0^{0i}\propto q_i$ and therefore $\chi_0^{0j}\left(\hat{P}_T\right)_{ji}=0$, the anisotropic density-current function acquires a finite {\em transverse} component:
\be
\lb{dcani}
\Pi_0^{0i}(q)=\frac{i\O_n q_i}{|\bq|^2}
\Pi_0^{00}(q)+\Pi_0^{0j}(q)\left(\hat{P}_T(\bq)\right)_{ji}.
\ee
Analogously, the current-current function $\Pi_0^{ij}$ does not admit a longitudinal-transverse decomposition as in Eq.\ \pref{cciso}, but it reads
\bea
\lb{ccani}
&&\Pi_0^{ij}(q)=
\frac{(i\O_n)^2}{|\bq|^2}
\Pi_0^{00}(q)
\left(\hat{P}_L(\bq)\right)^{ij}+
\Pi_0^T(q)
\left(\hat{P}_T(\bq)\right)^{ij}\nn\\
&&+\frac{i\O_n q_i}{|\bq|^2}
\Pi_0^{0k}(q)
\left(\hat{P}_T(\bq)\right)_{kj}
+\frac{i\O_n q_j}{|\bq|^2}
\Pi_0^{0k}(q)\left(\hat{P}_T(\bq)\right)_{ki},\nn\\
\eea
where $\Pi_0^T\equiv (1/2)\left(\hat{P}_T\right)_{ij}\Pi_0^{ji}$ is the purely transverse part. 
Eq.\ \pref{dcani} encodes the physical mechanism highlighted in Sec.\ \ref{retmax} within the formalism of Maxwell's equations, i.e. the possibility in a layered system to get a density fluctuation in response to a transverse current perturbation and viceversa.  From the point of view of the present derivation, the crucial consequence of Eq.\ \pref{dcani} is that the terms 
$\phi\Pi_0^{0i}A_i$, that now replace the corresponding ones of the isotropic case in Eq.\ \pref{sgb}-\pref{sgc}, are no more zero, even in the Coulomb gauge, leading to a finite coupling between internal (or external) scalar and vector potentials. In other words, the gauge-fixing procedure does not provide a decoupling between longitudinal and transverse degrees of freedom, as represented by the internal scalar and vector potentials, respectively. If one ignores the latter and retains only the former, as it is usually done in the context of RIXS and EELS experiments\cite{grecu_prb73,dassarma_prb82,tselis_prb84,lee_rixs_nature18,liu_rixs_npjqm20,zhou_prl20,huang_rixs_prb22,phillips_prb22}, one obtains the generalization of Eq.s \ \pref{stadrpadd}-\pref{stadrpacc} to the layered metal:
%
%
%From the point of view of the density response, that we are interested to discuss below, the main consequence is that while in the isotropic case it is only dressed by fluctuations of the internal scalar potential, corresponding to RPA resummation of the Coulomb interaction $V_C(\bq)$, in the layered case it is also dressed by fluctuations of the vector potential, carrying information on the current response. Since the coupling to the vector potential carries an additional $1/c$ prefactor, see Eq.\ \pref{sgaussint}, we expect that such "relativistic" corrections will only be relevant in a suitable low-momentum regime, as we anticipated in Sec.\ \ref{retmax} and as will be discussed in the next section. To better highlight the difference between the isotropic and anisotropic case we will express the full layered correlation functions $\tilde{\Pi}^{\mu\nu}$, as obtained by the simultaneous integration of the 4-potential $A_\mu$, as the sum of the standard-RPA contribution $\Pi_{RPA}^{\mu\nu}$ in the Coulomb interaction equivalent to Eqs. i.e.:
%
\be
\lb{stadrpaani}
\Pi_{RPA}^{00}(q)=
\frac{\Pi_0^{00}(q)}
{1-V_C(\bq)\Pi_0^{00}(q)},
\ee
\be
\lb{anirpadc}
\Pi_{RPA}^{0i}(q)=\frac{\Pi_0^{0i}(q)}
{1-V_C(\bq)\Pi_0^{00}(q)},
\ee
\be
\lb{anirpacc}
\Pi_{RPA}^{ij}(q)=\Pi_0^{ij}(q)
+V_C(\bq)
\frac{\Pi_0^{i0}(q)\Pi_0^{0j}(q)}
{1-V_C(\bq)\Pi_0^{00}(q)}.
\ee
The standard-RPA anisotropic functions defined above satisfy, as their isotropic counterparts, the gauge-invariance conditions
\be
\lb{girpaani}
q_\mu\Pi_{RPA}^{\mu\nu}(q)=0,\quad  \Pi_{RPA}^{\mu\nu}(q)q_\nu=0.
\ee
%
%%
%\be
%\lb{form}
%\tilde{\Pi}^{00}(q)=
%\Pi^{00}_{RPA}(q)+\Pi^{00}_{MIX}(q)
%\ee
%%
%
Also, Eqs.\ \pref{anirpadc} and \pref{anirpacc} can be put in the forms prescribed by Eqs.\ \pref{dcani} and \pref{ccani} for their bare counterparts, provided that one defines the transverse part of \pref{anirpacc} as $\Pi_{RPA}^T\equiv(1/2)\left(\hat{P}_T\right)_{ij}\Pi_{RPA}^{ji}$. In such "non-relativistic" limit the density-density and density-current response function of a layered systems are fully exhausted by Eq.s\ \pref{stadrpaani} and \pref{anirpadc} respectively. In particular, following the same reasoning of Eq.\ \pref{isodisp} above, the poles of Eq.\ \pref{stadrpaani} yield, in the long-wavelength dynamical limit in which $\Pi_0^{00}\simeq n/\o^2\left( q_{xy}^2/m_{xy}+q_z^2/m_z \right)$ ($q_{xy}\equiv\sqrt{q_x^2+q_y^2}$ being the in-plane momentum), the dispersion of the purely longitudinal layered plasmon usually quoted in the literature \cite{fetter_ap74,krezin_prb88,markiewicz_prb08,greco_cp19}, i.e.
\bea
\lb{laylong}
\o_L^2(\bq)&=&\o_{xy}^2\frac{q_{xy}^2}{|\bq|^2}+\o_z^2\frac{q_z^2}{|\bq|^2}\nn\\
&=&\o_{xy}^2\sin^2\eta + \o_z^2\cos^2\eta,
\eea
where $\eta$ denotes as before the angle between $\bq$ and the $z$-axis. In full analogy, one could define the dispersion of the anisotropic plasma polariton by the generalization of Eq.\ \pref{polar} to the anisotropic case, i.e. $1-D_T\Pi^T_0=0$. In this case, by exploiting the long-wavelength dynamical limit $\Pi_0^T\simeq n\left( q_{z}^2/m_{xy}+q_x^2/m_z \right)/|\bq|^2$ of the transverse anisotropic current response, one gets
\bea
\lb{laytrasv}
\o_T^2(\bq)&=&\o_{xy}^2\frac{q_z^2}{|\bq|^2}+\o_z^2\frac{q_{xy}^2}{|\bq|^2}+\tc^2|\bq|
\nn\\
&=&\o_{xy}^2\cos^2\eta + \o_z^2\sin^2\eta
+\tc^2|\bq|.
\eea
One can immediately notice that the expressions \pref{laylong}-\pref{laytrasv} are non-analytic functions as $\bq\ra \bo$. As such, they predict a continuum of possible values as $\bq\ra\bo$, with $\omega_T$ being even smaller than $\omega_L$ {at specific values of the angle $\eta$, leading to a crossing between the two dispersions as $\bq$ is increased}. {These features, in particular the continuum of $\bq=\bo$ values,} are unphysical, and in direct contrast with Maxwell's equation expectation, as we will discuss below. On the other hand, they provide a valid approximation for the non-relativistic limits {(i.e. when the coupling is negligible, see below)} of the generalized plasma modes valid in the anisotropic case at generic value of the wavevector.

In order to account for the finite coupling between the scalar and vector potentials we must {go back to Eq.\ \pref{sgaussint} in the anisotropic case (i.e. with $\chi\ra\Pi$) and integrate out the e.m. potentials $A_\mu$.} In the following, instead of choosing a particular gauge we will employ the so-called Faddeev-Popov gauge-fixing procedure\cite{nagaosa}, which consists in spoiling explicitly the gauge invariance of the model by adding a term
\be
\lb{gf}
S_{gf}[A_\mu]=
\frac{1}{2\a}\int d\t d\br 
\left[f\left(A_\mu\right)\right]^2
\ee
{that is not gauge-invariant.} In Eq.\ \pref{gf} $\a$ is an arbitrary multiplicative constant, and $f$ is a generic linear function of the 4-potential. $f^2$ is thus quadratic in $A_\mu$, and therefore Eq. \pref{gf} is a gaussian function of the 4-potential centered around the zeroes of $f(A_\mu)=0$. As a consequence, contributions from $A_\mu$ that do not satisfy such a condition are exponentially suppressed, {and one is guaranteed that, whilst the gauge-invariance of the physical quantities is preserved, the divergence associated with the infinite gauge orbits is eliminated.} A particular case is the $\a\ra 0$ one, i.e. when the width of the gaussian vanishes and only fields obeying exactly $f(A_\mu)=0$ survive in the functional integral. In our case, in order to reproduce the Coulomb gauge in the $\a\ra 0$ limit we choose $f(A_\mu)=\bdnb\cdot\bA$. The advantage of the Faddeev-Popov method lies in the fact that the potentials remain linearly independent, so that there is no need to parameterize them according to the chosen gauge and one can easily integrate out all the four component of $A_\mu$. If we fix the multiplicative constant as $\a=4\pi$ Eq. \pref{gf} becomes, in Fourier space,\\
\bea
\lb{gfq}
S_{gf}[A_\mu]&=&\frac{1}{8\pi}\sum_q
q_i q_j A_i(q)A_j(-q)\nn\\
&=&\frac{1}{8\pi}\sum_q
\left|\bq\cdot \bA(q)\right|^2,
\eea
which, as one immediately sees, cancels out with the $-1/(8\pi)\sum_q
\left|\bq\cdot \bA\right|^2$ coming from Eq. \pref{sem2}. The total action for the layered system will then be given by Eq. \pref{sgaussint}, without the last term of Eq.\ \pref{sgd}, once the $\chi$'s are replaced with the $\Pi$'s. We then integrate, as a first step, the scalar potential. This amounts to the standard-RPA dressing of the bare response function, plus a dressing of the remaining terms proportional to the vector potential:
\begin{widetext}
\bea
\lb{phiint}
S[\bA,A_\mu^{ext}]&=&\frac{e^2}{2}
\sum_q\left[\Pi_{RPA}^{00}
|\phi^{ext}(q)|^2+\Pi_{RPA}^{ij}(q) 
\frac{A_i^{ext}(q)}{c}
\frac{A_j^{ext}(-q)}{c}
\right.\nn\\
&+&\left(-
\Pi_{RPA}^{0i}(q)\phi_{ext}(q)
\frac{A_i^{ext}(-q)}{c}
-\Pi_{MIX}^{0i}(q)
\phi^{ext}(q)\frac{A_i(-q)}{c}
+\Pi_{MIX}^{ij}(q)
\frac{A_i^{ext}(q)}{c}\frac{A_j(-q)}{c}+c.c.\right)\nn\\
&+&\left.
\left(\L^{-1}(q)\right)^{ij}
\frac{A_i(q)}{c}\frac{A_j(-q)}{c}\right].\nn\\
\eea
The coefficient of $A_i A_j$ is defined as the inverse of the 3$\times$3 tensor $\L^{ij}$ as
\be
\lb{lambda}
(\L^{-1})^{ij}(q)=
\frac{c^2|\bq|^2}
{4\pi e^2}
\left(\hat{P}_L(\bq)\right)^{ij}+
\left(
\Pi_{RPA}^T(q)-\frac{1}{D_T(q)}\right)
\left(\hat{P}_T(\bq)\right)^{ij},
\ee
\end{widetext}
and the coefficients of the mixed terms in $\phi^{ext}A_i$ and in $A_i^{ext}A_j$ are given, respectively, by
\be
\lb{brpadc}
\Pi_{MIX}^{0i}(q)=\Pi_{RPA}^{0i}(q)
-\frac{i\O_n q_i}{|\bq|^2}\Pi_{RPA}^{00}(q),
\ee
\be
\lb{brpacc}
\Pi_{MIX}^{ij}(q)=\Pi_{RPA}^{ij}(q)
-\frac{i\O_n q_j}{|\bq|^2}\Pi_{RPA}^{0i}(q).
\ee
$\Pi_{MIX}^{0i}$ is purely transverse, since $\Pi_{MIX}^{0i}q_i=0$, while $\Pi_{MIX}^{ij}$, which is not symmetric under exchange of the indices, is such that $\Pi_{MIX}^{ij}q_j=0$ but $q_i\Pi_{MIX}^{ij}$ is finite. This is even clearer when one writes
\be
\lb{pioimix}
\Pi_{MIX}^{0i}(q)\equiv \Pi_{RPA}^{0j}(q)
\left(\hat P_T(\bq)\right)_{ji}
\ee
and
\bea
\lb{piijmix}
\Pi_{MIX}^{ij}(q)&\equiv&
\Pi_{RPA}^T(q)\left(\hat{P}_T(\bq)\right)^{ij}\nn\\
&+&\frac{i\O_n q_i}{|\bq|^2}
\Pi_{RPA}^{0k}(q)\left(\hat{P}_T(\bq)\right)_{kj},
\eea
which can be straightforwardly obtained by taking advantage of the fact that $\Pi_{RPA}^{0i}$ and $\Pi_{RPA}^{ij}$ admit two decompositions similar, respectively, to those of Eqs.\ \pref{dcani} and \pref{ccani}. Eq.s\ \pref{pioimix} and \pref{piijmix} clarify two main aspects. First of all, by direct comparison between Eqs.\ \pref{brpadc} and Eq.\ \pref{dcani} one sees that $\Pi_{MIX}^{0i}$ is exactly the transverse part of the standard-RPA density-current response function, that is not zero only in the layered case. Thus, the coupling between the scalar and vector potentials in Eq.\ \pref{phiint} is a direct consequence of the system anisotropy. At the same time, since $\Pi_{MIX}^{0i}q_i=0$ and $\Pi_{MIX}^{ij}q_j=0$, the external scalar $\phi^{ext}$ and vector $\bA^{ext}$ potentials in Eq.\ \pref{phiint}  couple only to the transverse part of the internal vector potential $\bA_T$. As a consequence, the gauge-dependent longitudinal component $\bA_L$, which only appears in the last term of Eq.\ \pref{phiint} via the quadratic contribution $(1/\L_L) |A_L|/c^2$, with $1/\L_L=c^2|\bq|^2/(4\pi e^2)$, does not contribute to the dressing of the response functions.  Once the integration of $\bA$ in Eq. \pref{phiint} is carried out  we are left with the action for the auxiliary fields only
\be
\lb{seffani}
S[A_\mu^{ext}]=\frac{e^2}{2 c^2}\sum_q A_\mu^{ext}(q)\tilde{\Pi}^{\mu\nu}(q)A_\nu^{ext}(-q).
\ee
The full response functions $\tilde{\Pi}^{\mu\nu}$ are given by
\bea
\lb{beyondrpa}
\tilde{\Pi}^{\mu\nu}(q)&=&\Pi_{RPA}^{\mu\nu}(q)-
\Pi_{MIX}^{\mu i}(q)
\L_{ij}^T(q)
\Pi_{MIX}^{\nu j}(q),\nn\\
\eea
where
\be
\L_{ij}^T\equiv\left(\hat{P}_T\right)_{ij}/\left( \Pi_{RPA}^T- 1/D_T\right)
\ee
and we used a compact 4-vector notation in which the mixing coefficients \pref{brpadc}-\pref{brpacc} are defined as
\be
\lb{4mix}
\Pi_{MIX}^{\mu i}(q)=\Pi_{RPA}^{\mu i}(q)-
\frac{i\O_n q_i}{|\bq|^2}\Pi_{RPA}^{0\mu}(q).
\ee
The full response functions \pref{beyondrpa} are still gauge invariant. Indeed, thanks to the gauge-invariance conditions \pref{piogi} and \pref{girpaani} for the bare and standard-RPA response functions respectively, we also have
\be
\lb{gianidress}
q_\mu\tilde{\Pi}^{\mu\nu}(q)=0,\quad
\tilde{\Pi}^{\mu\nu}(q)q_\nu=0
\ee
for the full response functions. 
To show that Eq. \pref{gianidress} is indeed satisfied, we note that each $\tilde{\Pi}^{\mu\nu}$ is given as the sum of a standard-RPA function $\Pi_{RPA}^{\mu\nu}$, which already obeys a gauge-invariant constraint, i.e. Eq. \pref{girpaani}, plus the mixing term \pref{4mix}, which can be proven to obey a similar condition. Indeed, from $q_\mu\Pi_{RPA}^{\mu \nu}=0$ and $\Pi_{RPA}^{\mu \nu}q_\nu=0$, it follows that
\bea
&& q_\mu\Pi_{MIX}^{\mu i}=
q_\mu\Pi_{RPA}^{\mu i}-
\frac{i\O_n q_i}{|\bq|^2}\Pi_{RPA}^{0\nu}q_\nu
=0,\nn\\
\eea
where the vanishing of the first and the second terms comes from Eq. \pref{gi} for $\nu=i$ and $\mu=0$ respectively. 

In the following we will be interested in the density-density function $\tilde\Pi^{00}$, whose expression is given by Eq. \pref{beyondrpa} for time-like indices $\mu=\nu=0$ and reads explicitly
\bea
\lb{beyondrpadd}
\tilde{\Pi}^{00}(q)&=&\Pi_{RPA}^{00}(q)-
\Pi_{MIX}^{0i}(q)
\L_{ij}^T(q)
\Pi_{MIX}^{0j}(q).\nn\\
\eea
Eq.\ \pref{beyondrpadd} is the first central result of our work. It provides an analytical expression for the gauge-invariant density-density response function in a layered metal at arbitrary momentum and frequency. It can be readily extended to the case of interacting electron systems, once that short-range interactions are included\cite{katsnelson_prl14,parcollet_prb14} by preserving the gauge-invariant condition \pref{piogi} for the bare response function. Indeed, all effects coming from the coupling to the long-range part of the interaction are included in an exhaustive way by Eq.\ \pref{beyondrpa}. In the next section we will provide additional analytical insights into the nature of the plasma modes obtained as poles of the general structure \pref{beyondrpadd} for the non-interacting case.

\section{Collective modes of a layered system and their spectral features}\lb{spec}

\subsection{Generalized plasma modes dispersion}
To analyze the spectral function of the density response  \pref{beyondrpadd} we set the momentum $\bq$ for the sake of simplicity within the $xz$ plane. With such choice the longitudinal-transverse basis is spanned by the following orthogonal vectors:
\be
\lb{ltbasis}
\hv_L=\hat{\bq},\quad
\hv_T^y=\hat{\by},\quad
\hv_T^{xz}=\hat{\bq}\times\hat{\by}.
\ee
$\hv_L$ is the longitudinal versor parallel to $\bq$, $\hv_T^y$ and $\hv_T^{xz}$ are the transverse ones along the $y$-direction and in the $xz$-plane respectively. To make a bridge between the longitudinal and transverse projectors defined in Eq.\ \pref{proj} and the basis versors \pref{ltbasis}, we notice that the former can be expressed in terms of outer products of the latter as $\hat{P}_L=\hv_L\otimes\hv_L$ and $\hat{P}_T=\hv_T^y\otimes\hv_T^y+\hv_T^{xz}\otimes\hv_T^{xz}$.\\
Let us now write the relevant layered response functions within the new basis \pref{ltbasis}. First of all, we note that the mixing terms \pref{brpadc} and \pref{brpacc} couple longitudinal excitations to transverse ones polarized along the $xz$ plane but not with those polarized along $y$. Indeed, since in our frame of reference $q_y=0$, we have that $\Pi_0^{0y}=0$ (which follows trivially from Eq.\ \pref{pio}) and, as a consequence, $\Pi_{MIX}^{0y}=0$. Therefore
\be
(\hv_T^y)_i\Pi_{MIX}^{0i}(q)=0,
\ee
Similarly Eq.\ \pref{pij}, for $i=y$ or $j=y$, has only a diamagnetic contribution proportional to a delta $\d_{yj}$ or $\d_{iy}$, so that
\be
(\hv_T^y)_i(\hv_L)_j\Pi_{MIX}^{ij}(q)=0,
\quad
(\hv_T^y)_i(\hv_T^{xz})_j\Pi_{MIX}^{ij}(q)=0
\ee 
which implies that current fluctuations along $y$ never get coupled with those along the longitudinal and the transverse $xz$ directions. As a consequence, the current-current response function, as given by Eq.\ \pref{beyondrpa} for space-like indices $\mu=i$ and $\nu=j$, reads, for $i=j=y$,
\be
\lb{currenty}
\tilde{\Pi}^{yy}(q)=\frac{\Pi_0^{yy}(q)}
{ 1-D_T(q)\Pi_0^{yy}(q) }.
\ee
Eq.\ \pref{currenty} identifies a transverse mode with electric field polarized along $y$ and whose long-wavelength propagation is given (being $\Pi_0^{yy}\simeq n/m_{xy}$ in the dynamical long-wavelength limit) by $\o^2=\o_{xy}^2+\tc^2|\bq|^2$, which coincides with the standard polariton dispersion \pref{polar}.

Conversely, Eqs.\ \pref{pioimix} and \pref{piijmix} account for a finite coupling between longitudinal modes and transverse ones polarized along $xz$. To investigate this effect it is convenient to express Eq.\ \pref{beyondrpadd} as
%
%\begin{widetext}
%\be
%\lb{Pioo}
%\tilde{\Pi}^{00}(q)=
%\frac{1}{V_C(\bq)}
%\left[
%\frac{D_T^{-1}(q)
%-\Pi_T^{JJ}(q)
%}
%{\left(
%1-V_C(\bq)\Pi_0^{00}(q)
%\right)
%\left(D_T^{-1}(q)
%-\Pi_T^{JJ}(q)
%\right)-
%V_C(\bq)
%\left(\Pi_T^{0J}(q)\right)^2}-1
%\right],
%\ee
%\end{widetext}
%
%
\be
\lb{Pioo}
\tilde{\Pi}^{00}(q)=
\frac{{\Pi}_{ret}^{00}(q)}{1-V_C(\bq){\Pi}_{ret}^{00}(q)},
\ee
where
\be
\lb{tildePioo}
{\Pi}_{ret}^{00}(q)=\Pi_0^{00}(q)+
\frac{\left(\Pi_T^{0J}(q)\right)^2}{ D_T^{-1}(q) - \Pi_T^{JJ}(q) }
\ee
and we introduced the projections of the bare density-current and current-current functions along the transverse $xz$ direction:
\bea
\lb{PiTOJJJ}
\Pi_T^{0J}(q) &\equiv& 
(\hv_T^{xz})_i 
\Pi_0^{0i}(q),\nn\\
\Pi_T^{JJ}(q) &\equiv& 
(\hv_T^{xz})_i 
(\hv_T^{xz})_j 
\Pi_0^{ij}(q).
\eea
Eq.\ \pref{Pioo}, whose derivation is detailed in Appendix \ref{appc}, is the second central result of this work, as it provides an expression for the full density-density response of the layered metal in terms of the bare anisotropic response functions, with the inclusion of the retardation effects. {It is important to stress that both Eq.\ \pref{beyondrpa} and Eq.\ \pref{Pioo} emphasize two main aspects: the former underlines that the anisotropy mixes the standard-RPA result, given by the integration of $\phi$, with the propagating transverse modes encoded into $\bA$; the latter suggests that this is equivalent to resuming first the retarded interaction mediated by the transverse modes and then the Coulomb interaction.} Indeed, according to Eqs.\ \pref{Pioo} and \pref{tildePioo}, the full density response contains an RPA resummation both in the current and density sector: the former accounts for the replacement of the bare bubble $\Pi_0^{00}$ with the correction due to the transverse gauge field, whose propagator is proportional to  $1/\left(D_T^{-1}-\Pi_T^{JJ}\right)$, with a strength controlled by the non-zero value of $\Pi_T^{0J}$ for a layered system; the latter is the usual RPA dressing with the Coulomb interaction $V_C(\bq)$.

We now focus, in full analogy with Sec.\ \ref{iso}, on the simplified long-wavelength dynamical case in which the anisotropic Lindhard bubbles \pref{poo}-\pref{pij} can be expanded, at leading order in the momentum, as 
%
%Moreover, at $|\bq|<q_c$ and $\d\ll\o_{xy/z}$, we can take advantage of the  the bare functions $\Pi_0^{00}$, $\Pi_T^{0J}$ and $\Pi_T^{JJ}$ in order to compute \pref{Pioo} within the same regime. In general, the expansion of the bare anisotropic response functions, i.e. Eqs.\ \pref{poo}, \pref{pio} and \pref{pij} within such a regime can be computed by taking advantage of the analogy between the layered electron gas and the isotropic one with mass $m^*=(m_{xy}^2 m_z)^{1/3}$, as prescribed by the aforementioned equations. In this case the long-wavelength dynamical limit can be identified either in terms of the rescaled mass and momentum $m^*$ and $\tilde{\bq}$ as $v_F^*|\tilde{\bq}|\ll\o$, where the auxiliary Fermi velocity $v_F^*$ is defined as $v_F^*\equiv\sqrt{2\e_F/m^*}$, or in terms of the parameter of the layered system as $v_F^{xy}|\bq|\ll\o$, where $v_F^{xy/z}\equiv\sqrt{2\e_F/m_{xy/z}}$ (with $v_F^{xy}\gg v_F^z$). From Eqs.\ \pref{approxt} and \pref{approx0broad} we have that $\chi_{0*}^T\approx n/m^*$, which as usual is not affected by the broadening at leading orders in $v_F^*|\tilde{\bq}|/\o$, and $\chi_{0*}^{00}(\tilde{\bq},\o)\approx n|\tilde{\bq}|/(m\o^2)\left(1-2i\d/\o\right)$. Substituting these two expansions into Eqs.\ \pref{poo}-\pref{pij} we finally have that:
%
\bea
\lb{piooo}
\Pi_0^{00}(q)&\simeq&
\frac{n}{\o^2}
\left(\frac{q_x^2}{m_{xy}}+
\frac{q_z^2}{m_z}\right),\\
\lb{piojt}
\Pi_{T}^{0J}(q)&\simeq&
-\frac{q_x q_z}{|\bq|}
\frac{n}{\o}
\left(\frac{1}{m_{xy}}-\frac{1}{m_z}\right),\\
\lb{pijjt}
\Pi_{T}^{JJ}(q)&\simeq&
\frac{n}{|\bq|^2}
\left(\frac{q_x^2}{m_z}+\frac{q_z^2}{m_{xy}}\right).\nn\\
\eea
The derivation of these limits is discussed in Appendix \ref{appb}. By using Eqs.\ \pref{piooo}-\pref{pijjt} we can immediately find an estimate of the retardation corrections to the density response encoded in Eq.\ \pref{tildePioo}, in the same spirit of the discussion in Sec.\ \ref{retmax}. Indeed, we see that the relative correction to the density bubble $\Pi_0^{00}$ encoded into Eq.\ \pref{tildePioo} can be expressed, by means of the expansions \pref{piooo}-\pref{pijjt}, as
%
%\bea
%\Pi_{T}^{0J}(q)&\simeq&
%-\frac{q_x q_z}{4\pi e^2|\bq|\omega}
%\left(\omega^2_{xy}-\omega^2_z\right)
%\nn\\
%&\simeq& \frac{1}{V_C(\bq)}\frac{|\bq|}{\o}\frac{\tc^2 q_c^2}{|\bq|^2}
%\frac{\sin(2\eta)}{2}
%\eea
%
%
\begin{widetext}
\be
\lb{lim}
\frac{\left(\Pi_{T}^{0J}(q)\right)^2}
{\Pi^{00}_0(q)( D_T^{-1}(q) - \Pi_{T}^{JJ}(q)) }
\simeq
\frac{\left(\o_{xy}^2-\o_z^2\right)^2}
{\o_L^2\left(\o^2-\o_T^2(\bq)\right)}
\frac{q_x^2 q_z^2}{|\bq|^4}
\equiv
\frac{\tc^4 q_c^4}
{\o_L^2(\bq)\left(\o^2-\o_T^2(\bq)\right)}
\frac{\sin^2(2\eta)}{4}
\ee
\end{widetext}
where
\be
q_c\equiv\frac{\sqrt{\o_{xy}^2-\o_z^2}}{\tc}
\ee
as already defined (for $\e=1$) in Eq.\ \pref{cross} above, and with the definitions \pref{laylong} and \pref{laytrasv} of $\o_L$ and $\o_T$. Outside the light cone, as it is the case for EELS and RIXS,  momenta are such that $|\bq|\gg\o/c$. In this regime the term $\tc^2 |\bq|^2$ in $\omega_T(\bq)$ dominates in the denominator of Eq.\ \pref{lim}, while  $\tc q_c \simeq \o_{xy}$ is comparable to $\o_L$. {One then recovers the same scaling {condition} $\sim q_c^2/q^2\ll 1$ of Eq.\ \pref{cross} for the quantitative irrelevance of retardation effects.} Conversely, for experiments with THz light where $\o\simeq \o_z$ and $q=\o_z/\tc$, which is far smaller than the crossover value $q_c\sim\o_{xy}/\tc$, one sees that the denominator of Eq.\ \pref{lim} scales at leading order with $\o_z^2 (\tc q_c)^2$:
\bea
\omega_L^2(\omega_T^2-\omega_z^2)&=&(\omega_z^2+(\tilde c q_c)^2\cos^2\eta)\nn\\
&\times&(\omega_z^2+(\tilde c q_c)^2\sin^2\eta)\simeq\omega_z^2 (\tilde c q_c)^2.
\eea
When  replaced into Eq.\ \pref{lim} one finds again an overall factor scaling as $ (c q_c/\omega_z)^2=(q_c/q)^2 \gg 1$, and one recovers that relativistic corrections become crucial. In the following we will see how the above estimate reflects in the crossover from the relativistic to the standard RPA regime for the response function. 
\begin{figure*}[ht] 
    \centering
\includegraphics[width=\textwidth]{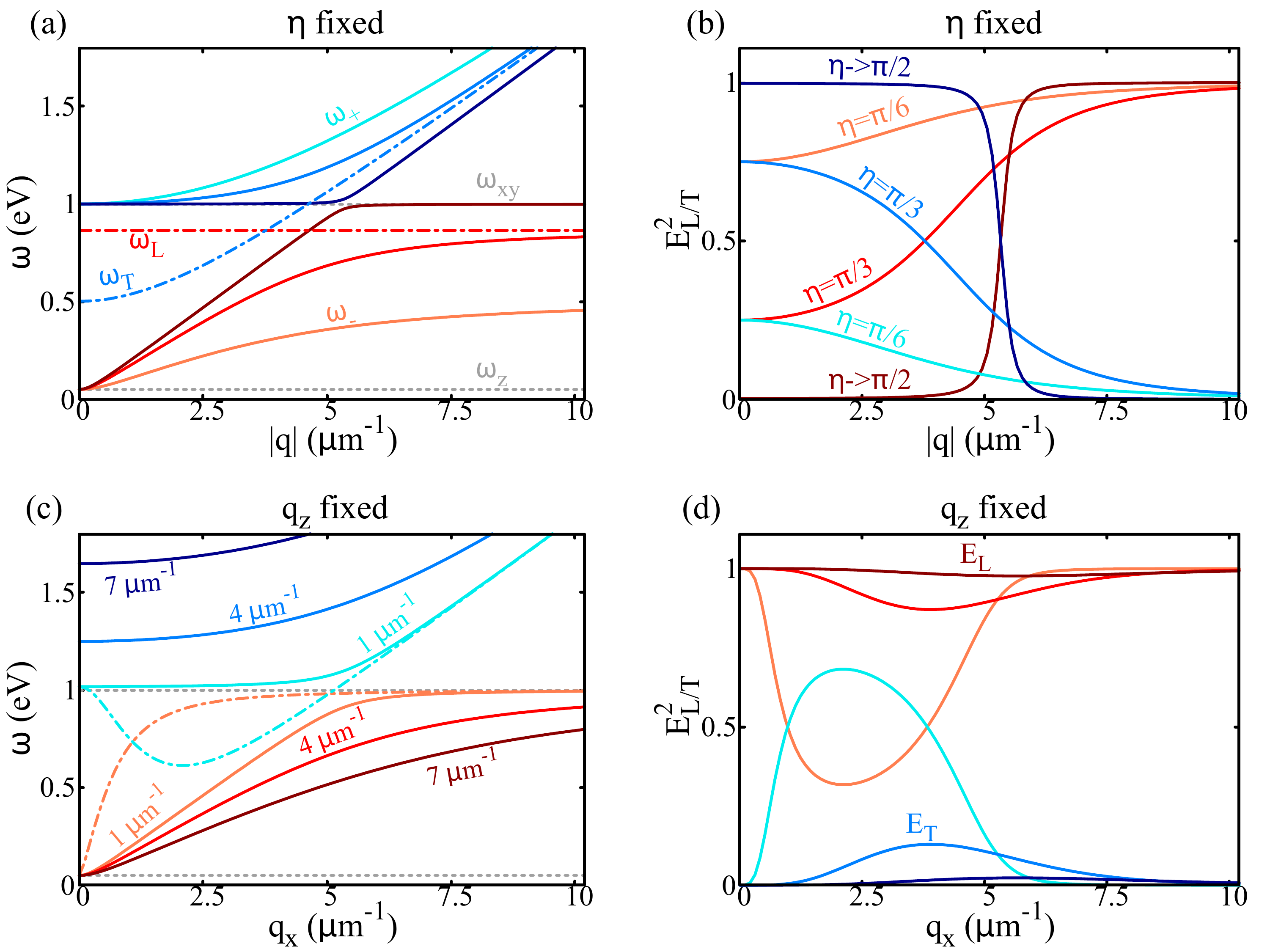}
\caption{{\bf Momentum dependence of the mixed longitudinal-transverse modes and their polarizations.} Momentum dependence of the mixed longitudinal-transverse modes $\o_-$ and $\o_+$ (panel (a) and (c)), as given by Eq.\ \pref{ltmodes}, and of the longitudinal and transverse components $E_L$ and $E_T$ of the $\omega_-$ mode, as given by Eqs.\ \pref{poll} and \pref{polt} (panel (b) and (d)). In panel (a) and (b) the quantities are shown as functions of $|\bq|$ at three selected values of the angle $\eta$, while in panel (c) and (d) they are shown as functions of $q_x$ at three selected values of the out-of-plane momentum $q_z$. Here red/blue solid lines are associated with $\o_-$/$\o_+$ and $E_L/E_T$, respectively, with same shades of red or blue associated with the same value of $\eta$ or $q_z$. In panel (a) and (c) we also show, for comparison, the standard-RPA modes $\o_L$ (red dot-dashed line) and $\o_T$ (blue dot-dashed line) at $\eta=\pi/3$  in panel (a) and $q_z=1$ $\m$m$^{-1}$ in panel (c), respectively. We set the values of the plasma frequencies as $\o_{xy}=1$ eV and $\o_z=0.05$ eV and for simplicity we assumed $\e=1$.}
\lb{fig1}
\end{figure*}

Let us first  determine the general dispersion of the plasma modes from the zeros of  Eq.\ \pref{Pioo}. With lengthly but straightforward calculations one obtains the expressions equivalent to those derived recently in Ref.\ \cite{gabriele_prr22} for an anisotropic superconductor in the case of zero screening length, i.e.
\bea
\lb{ltmodes}
&&\o_{\pm}^2(\bq)=\frac{1}{2}\left(
\o_{xy}^2+\o_z^2+\ct^2|\bq|^2\right.\nn\\
&\pm &\left. \sqrt{
(\o_{xy}^2-\o_z^2)^2+\ct^4|\bq|^4-2 \ct^2(q_x^2-q_z^2)(\o_{xy}^2-\o_z^2)
}
\right).\nn\\
\eea
As already discussed in Ref.\ \cite{gabriele_prr22}, in contrast to the RPA solution $\omega_{L/T}(\bq)$ of Eq.\ \pref{laylong}-\pref{laytrasv}, the two solutions $\omega_{\pm}$ are analytic functions as $\bq\ra \bo$, with
\be
\lim_{\bq\ra\bo} \omega_\pm(\bq)=\omega_{xy/z},
\ee
which is consistent with the expectation that as $\bq\ra\bo$ the only solution of the Ampere-Maxwell law $4\pi \bJ-(i\omega/c)\bE=0$ requires $\bJ$ being parallel to $\bE$, that is only possible in a layered system for electric fields polarized along $z$ and $\omega=\omega_z$, or polarized in the $xy$ plane with $\omega=\omega_{xy}$. Since at small but finite $\bq$ the two e.m. modes \pref{ltmodes} preserve the same polarization, for a generic direction of $\bq$ one has a mixture of longitudinal and transverse components. This can be explicitly seen by computing the polarization of the electric fields $\bE_-$ and $\bE_+$ corresponding to the two solutions $\omega_{\mp}$. Their expressions have been derived previously in the superconducting case\cite{gabriele_prr22}, and they can be obtained again within the framework of Maxwell's equations (see Appendix \ref{appa}). By denoting with $E_{L/T}$ the longitudinal/transverse component of the $\omega_-$ solution one finds\\
\be
\lb{freqmodes}
\begin{cases}
& \bE_-(\bq)=E_L(\bq)\hv_L(\bq)+
E_T(\bq)\hv_T^{xz},
\\
\\
&\bE_+(\bq)=E_L(\bq)\hv_T^{xz}-
E_T(\bq)\hv_L(\bq),
\end{cases}
\ee
where:
\be
E_L(\bq)=\frac{\frac{q_x q_z}{|\bq|^2}\left(\o^2_{xy}-\o_z^2\right)}
{\sqrt{\frac{q_x^2 q_z^2}{|\bq|^4}
\left(\o^2_{xy}-\o_z^2\right)^2+
\left(\o^2_+(\bq)-\o^2_T(\bq)\right)^2}},
\lb{poll}
\ee
\be
E_T(\bq)=
\frac{\o^2_+(\bq)-\o^2_T(\bq)}
{\sqrt{\frac{q_x^2 q_z^2}{|\bq|^4}
\left(\o^2_{xy}-\o_z^2\right)^2+
\left(\o^2_+(\bq)-\o^2_T(\bq)\right)^2}}.
\lb{polt}
\ee
As expected, for $\bE_+$ the role of $E_{L/T}$ is exchanged. The polarization vectors are shown along with the eigenmodes in Fig. \ref{fig1}(a)-(b) as functions of $|\bq|$ at fixed propagation angle $\eta$, and in Fig. \ref{fig1}(c)-(d) as functions of $q_x$ at fixed value of $q_z$. In both cases, one sees that as soon as $|\bq|\gg q_c$ the generalized modes tend to their RPA counterparts:
\be
\lb{nonrel}
|\bq|\gg q_c\Longrightarrow
\o_-(\bq)\simeq
\o_L(\bq),\quad
\o_+(\bq)\simeq
\o_T(\bq).
\ee
This can be easily understood from Eq.\ \pref{ltmodes}, where at large $|\bq|$ the square-root term can be recast and expanded in powers of the small variable $q_c/|\bq|$, and one easily recovers the two analytical expressions of the standard-RPA purely longitudinal and transverse modes \pref{laylong} and \pref{laytrasv}. In full agreement with this result, one sees that in such a non-relativistic regime $E_L\simeq 1$ and $E_T\simeq 0$, so that  $\bE_-$ describes a purely longitudinal mode while $\bE_+$ reduces to the purely transverse mode along $\hv_T^{xz}$, associated with the standard-RPA mode $\o_T$. In Fig.\ \ref{fig1}(a) and (c) we also show for comparison the standard-RPA results, for angle $\eta=\pi/3$ in Fig.\ \ref{fig1}(a) and for $q_z=1.5$ $\m$m$^{-1}$ in Fig.\ \ref{fig1}(c). As one can see, at small momenta the RPA solutions leads to the unphysical crossing of the transverse and longitudinal solutions, the former being even smaller in energy than the latter. Such a pathological behavior, that can be understood by the strong mixing of L/T character in the same regime of momenta, see Fig.\ \ref{fig1}(b) and (d), is completely solved by considering the generalized solutions. 
\begin{figure*}[ht!] 
    \centering
\includegraphics[width=\textwidth]{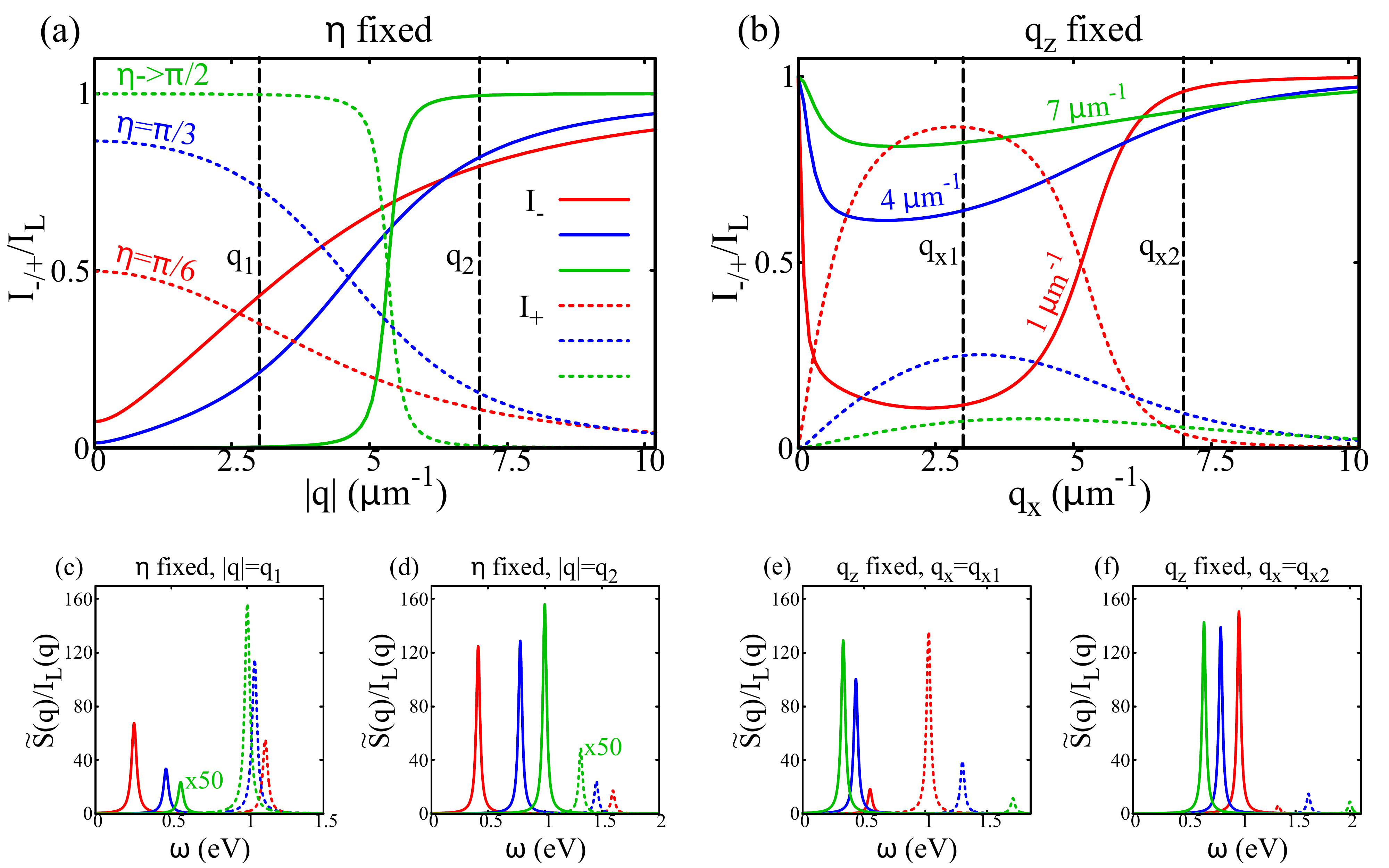}
\caption{{\bf Spectral features of the mixed longitudinal-transverse modes.} (a)-(b) Momentum dependence of the spectral weights $I_-$ and $I_+$, as given by Eq.\ \pref{ipm}, {both normalized with respect to the $I_L(\bq)$ defined in Eq.\ \pref{nnrelI-}.} In panel (a) the quantities are shown as functions of $|\bq|$ at three selected values of the angle $\eta$, while in panel (b) they are shown as functions of $q_x$ at three selected values of the out-of-plane momentum $q_z$. Here solid and dotted lines are associated with $I_-$ and $I_+$, respectively, with same color associated with the same value of $\eta$ or $q_z$. (c)-(f) Frequency dependence of the density response $\tilde{S}(q)$, as given by Eq.\ \pref{Srel}, at selected values of the momenta. In panels (c)-(d) we show the results at various angles $\eta$ for the two values $|\bq|=q_1$ and $q_2$  marked in in panels (a) and (b) by a vertical dashed line, with the same color convention used in (a). In panel (e)-(f) we show the results at different $q_z$ for the two values $q_x=q_{x,1}$ and $q_x=q_{x,2}$  marked in in panel (b) by a vertical dashed line, with the same color convention used in (b). In analogy with Fig.\ \ref{fig1} we set $\o_{xy}=1$ eV, $\o_z=0.05$ eV and $\e=1$. In order to better visualize the peaks associated with $\o_-$ and $\o_+$, we broadened the delta distributions with a finite width $\d=0.01$ eV. }
\lb{fig2}
\end{figure*}
%
 % Conversely In the relativistic regime $|\bq|<q_c$, as clear from the same figure, it can be $E_L>E_T$ or vice versa depending on the direction of $\bq$: at $\eta<\pi/4$, it is $E_L>E_T$ for all $\bq$, so that $\bE_-$ and $\bE_+$ always have, respectively, a predominant longitudinal and a predominant transverse character; at $\eta\geq\pi/4$ it is $E_L<E_T$ at very small $\bq$, $\bE_-$ and $\bE_+$ being thus predominantly transverse and longitudinal, with $E_L>E_T$ being recovered once $|\bq|>q_c f(\eta)$, where $f(\eta)\equiv\sqrt{\sin^2\eta-\cos^2\eta}$ is a positive and increasing function of the angle for $\pi/4\leq\eta\leq\pi/2$.\\

\subsection{Density response}

Once clarified the behavior of the generalized plasma modes let us study their contribution to the density response function $\tilde{S}(q)\equiv-\Im\tilde{\Pi}^{00}(q)$, whose general expression in terms of the retarded density function $\pref{tildePioo}$ is given by
\be
\lb{Sret}
\tilde{S}(q)=\frac{{-\Pi_{ret}^{00}}''(q)}
{\left(1-V_C(\bq){\Pi_{ret}^{00}}'(q)
\right)^2
+
\left( V_C(\bq)
{\Pi_{ret}^{00}}''(q) \right)^2
}.
\ee
The zeroes of $1-V_C{\Pi_{ret}^{00}}'$ identify the dispersions of the modes as given by Eq. \pref{ltmodes}. The imaginary part ${\Pi_{ret}^{00}}''$ sets the widths associated with their peaks, and it vanishes, as in the isotropic case, in the long-wavelength dynamical limit: in particular, it is ${\Pi_{ret}^{00}}''\ra 0^-$ for $|\bq|<q_c$ and $\o$ such that $\o=\o_{\pm}(\bq)$ (as we discuss at the end of Appendix \ref{appc}). In such a momentum range $\o_-$ and $\o_+$ are therefore well-defined modes and they both appear, in the density spectrum identified by $\tilde{S}$, as two sharp delta-like peaks centered around their respective frequencies, i.e.
\bea
\lb{Srel}
\tilde{S}(q)&\simeq&\pi I_-(\bq)\d\left(\o-\o_-(\bq)\right)\nn\\
&+&\pi I_+(\bq) \d\left(\o-\o_+(\bq)\right),
\eea
where the overall peak intensities are given by
\bea
\lb{ipm}
I_\pm(\bq)&=&\pm
\frac{\o_\pm(\bq)}{2 V_C(\bq)}
\frac{\o^2_\pm(\bq)-\o^2_T(\bq)}
{\o^2_+(\bq)-\o^2_-(\bq)}.
\eea
As a first observation, we notice that Eq.\ \pref{Srel} satisfies the $f$-sum rule, as one can prove by computing the integral $-(1/\pi)\int d\o\Im\tilde{\Pi}^{00}(\o,\bq)=(1/\pi)\int d\o\tilde{S}(\o,\bq)$ over all the positive frequencies. Taking advantage of the identity $\o_-^2+\o_+^2=\o_{xy}^2+\o_z^2+\tc^2|\bq|^2$, it yields
\bea
\lb{fsum}
&&\int_0^{+\infty}\frac{d\o}{\pi} 
\o\tilde{S}(\o,\bq)=\nn\\
&=&\o_+(\bq)I_+(\bq)+\o_-(\bq)I_-(\bq)
=\frac{n}{2}
\left(
\frac{q_x^2}{m_{xy}}+
\frac{q_z^2}{m_z}
\right),\nn\\
\eea
i.e. the expected result for the $f$-sum rule of an anisotropic electron gas \cite{vignale}.

The momentum dependence of the spectral weights is shown in Fig.\ \ref{fig2}(a) and Fig.\ \ref{fig2}(b) as a function of $|\bq|$ at fixed angle and as a function of $q_x$ at fixed $q_z$, respectively. As one can see, as $\bq\ra\bo$ in general the spectral function displays a two-peak structure, also shown explicitly in Fig. \ref{fig2}(c)-(f). This is a direct consequence of the fact, highlighted above, that at generic wavevector both modes have a finite longitudinal component in the relativistic regime. As a consequence, since the density response projects out the longitudinal fluctuations, it carries out a finite spectral weight at both modes. This is the third relevant result of the present work, that shows the emergence of a double-peak structure in the density spectral function in the relativistic regime. On the other hand, as the momentum increases and overcomes $q_c$ one sees from Eq. \pref{ipm} that
\be
\lb{nnrelI-}
I_+(\bq)\simeq 0,\quad I_-(\bq)\simeq I_L(\bq)\equiv\frac{\o_L(\bq)}{2 V_C(\bq)},\quad|\bq|\gg q_c
\ee
i.e. $I_+$ vanishes and  $I_-$ approaches the spectral weight expected for a standard-RPA longitudinal mode, i.e. the one given by Eq. \pref{Il} (see Fig. \pref{fig2}(a) and (c)) with $\o_L$ being now the standard-RPA plasmon defined in Eq.\ \pref{laylong}. Therefore, provided that $\bq$ does not exceed the value above which plasmons are damped {by the particle-hole continuum} (see Appendix \ref{appc}), one finds that\\
\be
\lb{mppeaknrel}
\tilde{S}(q)\simeq \pi I_L(\bq)
\d\left(\o-\o_L(\bq)\right),
\quad
|\bq|\gg q_c,
\ee
which is exactly the anisotropic counterpart of Eq.\ \pref{plasmawidth}. These effects are shown in Fig.\ \ref{fig2}(c)-(f), where we plot the spectral function $\tilde{S}$ as given by Eq. \pref{Srel} and normalized with respect to $I_L$, in order to get rid of the overall $\propto|\bq|^2$ factor due to charge conservation. $\tilde{S}/I_L$ is shown at the two values of the momentum $q_1=3.0$ $\m$m$^{-1}$ (below the crossover) and $q_2=7.0$ $\m$m$^{-1}$ (above the crossover) in (c), (e) and (d), (f) respectively: in the first case $\o_+$ carries on a larger spectral weight than $\o_-$, in agreement with the previous discussion on the behaviour of $I_\pm$ below $q_c$; in the second case $\o_-$ has the largest spectral weight, as its spectral profile of tends to the one of the standard-RPA plasmon, while $\o_+$ has an overall vanishing peak intensity, as expected for a pure polariton.

\section{Conclusions}\lb{conc}

In the present paper we provided a general derivation of the electromagnetic response functions of a layered electron gas in the long-wavelength limit that accounts for both instantaneous  and retarded electromagnetic interactions. The starting point is the observation that the anisotropy of the electric current induced as response to a local electric field implies, already at the level of classical Maxwell's equations, a mixing between the longitudinal and the transverse components of the internal e.m. fields. The physical effect is the emergence of a transverse current in response to a longitudinal electric field, that in turns acts as a source for the magnetic field. The final outcome is a mixing between the so-called polariton-like and plasmon-like hybrid light-matter modes, which appear instead in the isotropic metal as purely transverse and longitudinal modes, respectively. 

To implement this effect within a general many-body formalism we used a path-integral approach where the electronic degrees of freedom are explicitly coupled not only to the external (source) fields but also to the e.m. fields mediating the interactions among them. This approach highlights how the different role of the electric and magnetic field within the context of the Maxwell equation manifests in the usual language of the RPA resummation of the bare electronic response functions, that represent the standard paradigm to study plasma modes. The results can be summarized in the case e.g.  of the density response, that is the one probed by RIXS and EELS spectroscopy. In the isotropic system density fluctuations only couple to the scalar potential, and as a consequence at RPA level the density response is only dressed by Coulomb-like interactions. However, in the anisotropic system a transverse current is induced by a density fluctuations, leading to an additional RPA dressing of the density response via the transverse e.m. propagator. This is shown in Eq.\ \pref{Pioo} that we report here for convenience:
\be
\label{rpa}
\tilde{\Pi}^{00}(q)=
\frac{{\Pi}_{ret}^{00}(q)}{1-V_C(\bq){\Pi}_{ret}^{00}(q)}.
\ee
In Eq. \pref{rpa} the standard RPA resummation with the Coulomb potential is carried out using as starting point a density response function ${\Pi}_{ret}^{00}(q)$ that includes retardation effects, i.e.
\be
\lb{density}
{\Pi}_{ret}^{00}(q)=\Pi_0^{00}(q)+
\frac{\left(\Pi_T^{0J}(q)\right)^2}{ D_T^{-1}(q) - \Pi_T^{JJ}(q) }.
\ee
Retardation effects appear as a "relativistic" contribution, since one can show that they are negligible above the momentum threshold $q_c\sim\omega_{xy}/c\sim 5$ $\m$m$^{-1}$ vanishing for infinite light velocity.  Indeed,  at $q\gg q_c$ the second term of Eq.\ \pref{density} vanishes, and one recovers the textbook result \pref{rpa} with the standard bare density response $\Pi_0^{00}$, leading to a layered version of the longitudinal plasma mode. In contrast, in the low-momentum regime $q<q_c$ Eq.\ \pref{rpa} admits two poles, that coincide formally with the generalized plasma waves derived previously in Ref.\ \cite{gabriele_prr22}) in the superconducting state. The mixed longitudinal-transverse character of these modes manifests indeed as a finite projection of both modes in the density sector, leading to a double-peak structure of the density response. 
Such a prediction could be confirmed once that EELS and RIXS experiments will be able to push their resolution down to  the crossover scale. Indeed, despite the rather low state-of-the-art momentum resolution of these protocols, with the lowest accessible momentum of about $\sim 0.01$ \AA$^{-1}=100$ $\m$m$^{-1}$, electron energy loss spectroscopy incorporated in a scanning transmission electron microscope (STEM-EELS) and equipped with a monochromator and aberration correctors has a high potential to combine high momentum and energy resolution \cite{pichler_nature19,abajo_natmat21}, and thus to explore plasma excitations around the crossover scale, where the standard RPA breaks down and both generalized plasma modes give a comparable contribution to the density response. 

The main advantage of the derivation presented in this manuscript, encoded in a very compact and elegant way into Eq.\ \pref{density}, is the possibility to provide a general framework to study charged plasmon in a layered system by making explicit the effect of all long-range e.m. interactions, and leaving as a separate problem the inclusion of short-range interactions in the response functions $\Pi^{\mu\nu}$ that appear as building block of the final observable. The latter has been instead the focus of recent experiments of reflection EELS\cite{abbamonte_scipost17,mitrano_pnas18,mitrano_prx19,abbamonte_cm23}
in cuprates. In these materials an anomalous damping of plasmons occurs already at low momenta where particle-hole excitations are not operative, according to the standard Fermi-liquid description. Such a result has been attributed to a strange-metal physics\cite{abbamonte_natcomm23}, that is not captured by our approach, but can be in principle incorporated into the general expression \pref{density} by means of a proper gauge-invariant renormalization of the bare response functions due to short-range interactions. An additional interesting open question is the possibility that short-range interactions affect also the crossover scale, making relativistic effects operative below the momenta estimated on the basis of a Fermi-liquid picture. Such a mechanism could help the spectroscopic detection of the predicted generalized plasma modes, adding an additional knob to the investigation of electronic excitations in correlated metals.

\vspace{1cm} {\bf Acknowledgments}
We acknowledge financial support by EU under project MORE-TEM ERC-SYN (grant agreement No 951215) and by Sapienza University under the program Ateneo (No 2021 RM12117A4A7FD11B and 2022 RP1221816662A977).

\appendix

\section{Classical electrodynamics of a layered metal}\lb{appa}

In this appendix we rephrase the existence of mixed longitudinal-transverse e.m. modes in a layered metal within the classical framework of Maxwell's equations. A similar approach has been previously discussed for layered SC systems by three of us in Ref. \cite{gabriele_prr22}.

For the sake of simplicity, we consider a metal in the absence of external sources, i.e. $\rho_{ext}=0$ and $\bJ_{ext}=\bo$. The electron transport can be described, in the simplified case of long-wavelength propagation of the e.m. modes at very low scattering rate, by means of the undamped Drude equation for the internal current and electric field $\bJ$ and $\bE$:
\be
\lb{londeq}
\frac{\pd \bJ}{\pd t}=e^2 n \hat{m}^{-1} \bE.
\ee
$\hat{m}$ is the effective-mass tensor, that in isotropic systems trivially reduces to the scalar mass $m$ along an arbitrary direction; on the other hand, in layered anisotropic systems it reads 
\be
\hat{m}=\begin{pmatrix} m_{xy} & 0 & 0 \\ 0 & m_{xy} & 0 \\ 0 & 0 & m_z \end{pmatrix},
\ee
where $m_{xy}$ and $m_z$ are the in-plane and the out-of-plane effective masses, respectively. As it is usually done to derive the wave equation from the Maxwell's ones, one can take the curl of Faraday's law and then replace $\bB$ from Amp\'ere-Maxwell's equation. This yields the following equation for the electric field \cite{griffiths}:
\be
\lb{biosav}
\bdnb\left(\bdnb\cdot\bE\right)-\nb^2\bE=
\frac{4\pi}{c^2}\frac{\pd \bJ}{\pd t}-
\frac{\e}{c^2}\frac{\pd^2 \bE}{\pd t^2}
\ee
By exploiting Eq.\ \pref{londeq}, we get rid of $\bJ$ and obtain an equation for the electric field only. Let us introduce, as in the main text, the longitudinal $\bE_L=(\hat{\bq}\cdot\bE)\hat{\bq}$ and the transverse $\bE_T=\bE-\bE_L=(\hat{\bq}\times\bE)\times\hat{\bq}$ components of the electric field. In the isotropic case the longitudinal-transverse decomposition $\bE=\bE_L+\bE_T$ of the total electric field leads to two decoupled equations, i.e.
\be
\frac{\pd^2 \bE_L}{\pd t^2}+\o_p^2\bE_L=\bo,
\ee
\be
\frac{1}{\tc^2}
\frac{\pd^2 \bE_T}{\pd t^2}- \nb^2\bE_T+
\frac{\o_p^2}{\tc^2}
\bE_T=\bo,
\ee
where the renormalized light velocity is defined as $\ct=c/\sqrt{\e}$ as in the main text. They describe a longitudinal mode oscillating at $\o=\o_p$ and two degenerate transverse modes propagating at $\o^2=\o_p^2+\ct^2|\bq|^2$, $\o_p$ being the isotropic plasma frequency defined in Eq.\ \pref{pf}.

In the anisotropic case such a decomposition for the electric field does not decouple the two equations. The main physical reason is that, due to the tensorial nature of the effective mass, the induced current ${\bf J}$ in Eq.\ \pref{londeq} is no more parallel to the electric field. Let $\hat{x}$ be, as in the main text, the versor parallel to the direction of the in-plane component of the momentum $\bq$. For an anisotropic system Eq.\ \pref{biosav} splits into three equations. One of them describes the in-plane pure transverse component $\bE_T^y=E_T^{y}\hat{\by}$ through
\be
\lb{MaxY}
\left(\o^2-\o_{xy}^2-\ct^2|\bq|^2\right)E_T^y=0.
\ee
Such transverse mode, which is polarized along the $xy$ plane, is not affected by the anisotropy along the out-of-plane direction, so it propagates at $\o^2=\o_{xy}^2+\tc^2|\bq|^2$ without coupling with the longitudinal degrees of freedom. This is the result we found above with  Eq.\ \pref{currenty}. On the other hand, the two equations describing the longitudinal mode $\bE_L=E_L\hat{\bq}$ and the transverse component $\bE_T^{xz}=E_T^{xz}(\hat{\bq}\times\hat{\by})$ polarized along the $xz$ plane are coupled. Such equations read, in Fourier space:
\begin{widetext}
\be
\lb{MaxAni}
\begin{cases}
&\left(\o^2-\o_{xy}^2\frac{q_x^2}{|\bq|^2}-\o_z^2\frac{q_z^2}{|\bq|^2}\right)E_L
+\frac{q_x q_z}{|\bq|^2}\left(\o_{xy}^2-\o_z^2\right)E_T^{xz}=0,
\\
\\
&\left(\o^2-\o_z^2\frac{q_x^2}{|\bq|^2}-\o_{xy}^2\frac{q_z^2}{|\bq|^2}-\ct^2|\bq|^2\right)E_T^{xz}+
\frac{q_x q_z}{|\bq|^2}\left(\o_{xy}^2-\o_z^2\right)E_L=0.
\end{cases}
\ee
The non-trivial propagating solutions of the previous equations are found by solving the characteristic polynomial
\be
\lb{chareq}
\left(\o^2-\o_{xy}^2\right)
\left(\o^2-\o_z^2\right)
-\ct^2 q_x^2\left(\o^2-\o_{xy}^2\right)
-\ct^2 q_z^2\left(\o^2-\o_z^2\right)=0
\ee
\end{widetext}
that leads to the frequencies $\o_{\pm}$ introduced in Eq.\ \pref{ltmodes}. The electric fields $\bE_{\pm}$ associated with such modes can be then computed: they are given by Eq.\ \pref{freqmodes}. Notice that if the coupling term $\frac{q_x q_z}{|\bq|^2}(\o_{xy}^2-\o_z^2)E_T^{xz}$ is neglected in the first equation in Eq.\ \pref{MaxAni}, the pure longitudinal standard RPA mode $\o_L=\sqrt{\left(\o_{xy}^2 q_{xy}^2+\o_z^2 q_z^2\right)/|\bq|^2}$ is recovered. This is valid when the transverse component $E_T^{xz}$ is negligible with respect to the longitudinal one $E_L$, as expected when $|\bq|\gg q_c$. Indeed, from the second of Eq.\ \pref{MaxAni} one can estimate their ratio, at generic frequency and momentum, as
\be
\lb{ratioapp}
\frac{E_T^{xz}}{E_L}=-
\frac{\o_{xy}^2-\o_z^2}{\o^2-\o_T^2(\bq)}
\frac{q_x q_z}{|\bq|^2}=
-\frac{\tc^2 q_c^2}{\o^2-\o_T^2}
\frac{\sin\left(2\eta\right)}{2},
\ee
where $\o_T=\sqrt{\left(\o_{xy}^2 q_z^2 + \o_z^2 q_{xy}^2\right)/|\bq|^2+\tc^2|\bq|^2}$ is the standard-RPA pure transverse mode. Eq.\ \pref{ratioapp} is a more refined version of Eq.\ \pref{ratiolt}, where the displacement current was neglected, and is very similar to the one derived within the Many-Body formalism, i.e. Eq.\ \pref{lim}. As for the latter, when the momentum lies outside the light cone, i.e. $|\bq|\gg \o/\tc$, $E_T^{xz}/E_L\simeq 0$, while the ratio stays finite for THz light propagating with a wavevector $q=\o_z/c$ that lies far below the crossover value $q_c\sim \o_{xy}/\tc$.

\section{Relation between isotropic and anisotropic bare response functions}\lb{appb}

In this appendix we discuss the mapping of an anisotropic free-electron gas model into an isotropic one, i.e., strictly speaking, how to obtain the layered bare Lindhard functions $\Pi_0^{\mu\nu}$ from the knowledge of the isotropic ones $\chi_0^{\mu\nu}$. The former ones are given by an anisotropic generalization of Eqs.\ \pref{bare} and \pref{para} of the main text, i.e., after analytical continuation $i\O_m\ra\o+i 0^+$,
\bea
\lb{anilind}
&&\Pi^{\mu\nu}_0(\bq,\o)=
\frac{n}{m_i}\d^{\mu i}
\d^{\mu\nu}\left(1-\d^{\mu0}\right)\nn\\
&+& 2\int \frac{d^3 k}{(2\pi)^3} 
\tilde{\g}^{\mu}\left(\bk,\bq\right)
\tilde{\g}^{\nu}\left(\bk,\bq\right)
\frac{f(\tilde{\xi}_\bk)-f(\tilde{\xi}_{\bk+\bq})}
{\o-\tilde{\xi}_{\bk+\bq}+\tilde{\xi}_\bk+i 0^+}\nn\\
\eea
where we took the limit of infinite volume $V\ra\infty$. In Eq.\ \pref{anilind} $\tilde{\xi}_\bk\equiv k_{xy}^2/(2 m_{xy})+k_z^2/(2 m_z)-\mu$, with $k_{xy}\equiv\sqrt{k_x^2+k_y^2}$, is the anisotropic free-electron energy dispersion and $\tilde{\g}^\mu$ is the anisotropic density-current vertex, with $\tilde{\g}^0=1$ for $\mu=0$ and $\tilde{\g}^i=\left(k_i+q_i/2\right)/m_i$ for $\mu=i$. As a first step, we perform a rescaling of the momentum, in order to link e.g. the anisotropic energy dispersion to an isotropic one. In order to do so, we introduce the effective mass $m^*$, defined as
\be
m^*\equiv\left(
m_{xy}^2 m_z
\right)^{\frac{1}{3}}
\ee
and we perform the following rescaling of the momenta $\bk$ and $\bq$:
\be
\lb{scale}
k_i^*=\sqrt{\frac{m^*}{m_i}} k_i,\quad 
q_i^*=\sqrt{\frac{m^*}{m_i}} q_i.
\ee
Eq.\ \pref{scale} leaves the momentum integration measure invariant, i.e. $d^3 k^*=d^3 k$. It is straightforward to prove that the anisotropic energy dispersion and the current vertex can be rewritten, in terms of $m^*$, as
\be
\lb{xi}
\tilde{\xi}_\bk=
\frac{|\bk^*|^2}{2 m^*}-\mu
\equiv\xi_{\bk^*}^*
\ee
and
\be
\lb{g}
\tilde{\g}^i(\bk,\bq)=\sqrt{\frac{m^*}{m_i}}
\frac{k_i^*+q_i^*/2}{m^*}\equiv
\sqrt{\frac{m^*}{m_i}}
\g_*^i(\bk^*,\bq^*),
\ee
where $\xi^*$ and $\g^*$ are the energy dispersion and current vertex, as functions of the rescaled momenta \pref{scale}, of a fictitious isotropic free-electron gas with effective electron mass $m^*$.\\
Let us consider, as an example, the anisotropic density-density response function as given by Eq.\ \pref{anilind} for time-like indices $\mu=\nu=0$. Taking advantage of Eq.\ \pref{xi} and of the invariance of the integration measure under the rescaling \pref{scale} we have that
\bea
\lb{pooani}
\Pi_0^{00}(\bq,\o)&=&
2\int \frac{d^3 k}{(2\pi)^3}
\frac{f(\tilde{\xi}_\bk)-f(\tilde{\xi}_{\bk+\bq})}
{\o-\tilde{\xi}_{\bk+\bq}+\tilde{\xi}_\bk+i 0^+}\nn\\
&=&
2\int \frac{d^3 k^*}{(2\pi)^3}
\frac{f(\xi_{\bk^*}^*)-f(\xi_{\bk^*+\bq^*}^*)}
{\o-\xi_{\bk^*+\bq^*}^*+\xi_{\bk^*}^*
+i 0^+}\nn\\
&\equiv&\chi_{0*}^{00}(\bq^*,\o).
\eea
By means of similar calculations, one can show that the anisotropic density-current and current-current functions can be computed as
\bea
\lb{poiani}
\Pi_0^{0i}(\bq,\o)
&=&
\sqrt{\frac{m^*}{m_i}}\chi_{0*}^{0i}(\bq^*,\o)=\nn\\
&=&\sqrt{\frac{m^*}{m_i}}\frac{ \o q_i^* }{|\bq^*|^2}
\chi_{0*}^{00}(\bq^*,\o)
\eea
and
\bea
\lb{pijani}
\Pi_0^{ij}(\bq,\o)&=&
\sqrt{\frac{m^*}{m_i}}
\sqrt{\frac{m^*}{m_j}}\chi_{0*}^{ij}(\bq^*,\o)\nn\\
&=&\sqrt{\frac{m^*}{m_i}}
\sqrt{\frac{m^*}{m_j}}
\frac{\o^2}{|\bq^*|^2}
\chi_{0*}^{00}(\bq^*,\o)
\frac{ q_i^* q_j^*}
{|\bq^*|^2}+\nn\\
&+&\sqrt{\frac{m^*}{m_i}}
\sqrt{\frac{m^*}{m_j}}
\chi_{0*}^T(\bq^*,\o)
\left(\d_{ij}-
\frac{q_i^* q_j^*}
{|\bq^*|^2}\right).\nn\\
\eea
The last three equations are the same quoted in the main text, see e.g. Eqs. \pref{poo}, \pref{pio} and \pref{pij}. $\chi_{0*}^{00}$, $\chi_{0*}^{0i}$ and $\chi_{0*}^{ij}$ are, respectively, the bare density-density, density-current and current-current response functions of the isotropic free-electron gas with mass $m^*$, as functions of the rescaled momentum $\bq^*$ and of the frequency. Moreover, in the second row of Eqs.\ \pref{poiani} and \pref{pijani} we took advantage of the longitudinal-transverse decomposition, with respect to the momentum $\bq^*$, of $\chi_{0*}^{0i}$ and $\chi_{0*}^{ij}$, as prescribed for an isotropic metal by Eqs.\ \pref{dciso} and \pref{cciso} of the main text.
\begin{figure*}[ht!] 
    \centering
\includegraphics[width=\textwidth]{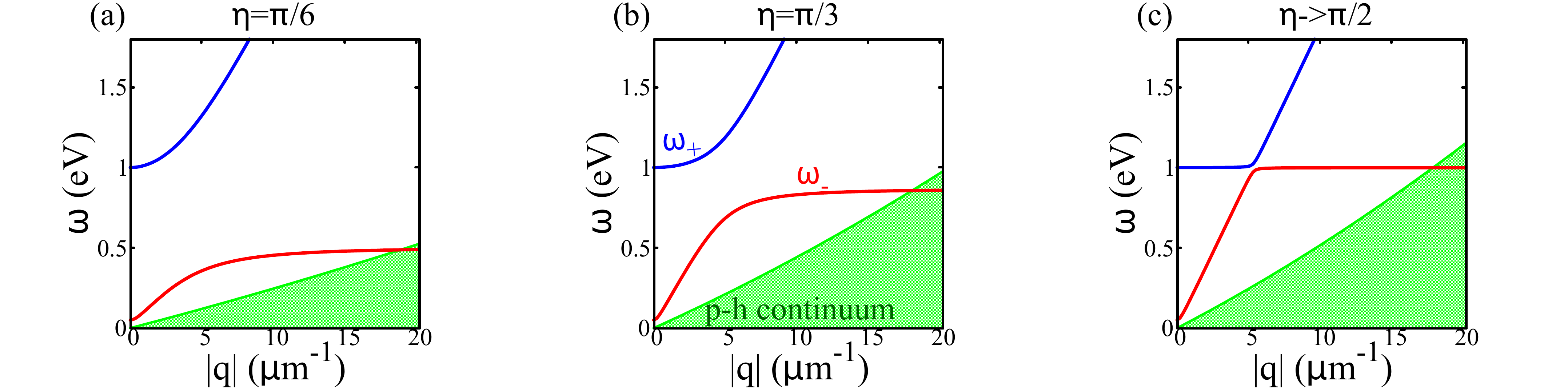}
\caption{{\bf Generalized plasma dispersions VS p-h continuum.} Momentum dependence of the mixed longitudinal-transverse modes $\o_-$ and $\o_+$ (red and blue solid lines respectively), as given by Eq.\ \pref{ltmodes} of the main text, and momentum and frequency dependence of the particle-hole (p-h) continuum (green-shaded area), as identified by the values of $\bq$ and $\o$ for which ${\Pi_{ret}^{00}}''\neq 0$. In all panels the quantities are shown as functions of $|\bq|$ at three selected values of the angle $\eta$. In analogy with the figures of the main text we set $\o_{xy}=1$ eV, $\o_z=0.05$ eV for the plasma frequencies and $\e=1$ for the background dielectric constant. Here we also fixed, in order to plot the p-h continuum, the value of the Fermi energy as $\e_F=1.0$ eV, that yields $v_F^{xy}\simeq 0.047$ eV$\m$m and $v_F^z\simeq 0.0024$ eV$\m$m for the Fermi velocities.}
\lb{fig3}
\end{figure*}

From Eqs.\ \pref{pooani}, \pref{poiani} and \pref{pijani} one easily derives the long-wavelength expansions \pref{piooo}, \pref{piojt} and \pref{pijjt} of the main text. Indeed, at leading order in $v_F^*|\bq^*|/\o$ ($v_F^*\equiv\sqrt{2 \e_F/m^*}$) it is $\chi_{0*}^{00}\simeq n|\bq^*|^2/(m^*\o^2)$ and $\chi_{0*}^T\simeq n/m^*$, in analogy with Eqs. \pref{approx0} and \pref{approxt} of the main text. Once substituted the definitions of $m^*$ and $\bq^*$ one finds that
\be
\Pi_0^{00}(\bq,\o)\simeq\frac{n}{\o^2}
\left(\frac{q_{xy}^2}{m_{xy}}+\frac{q_z^2}{m_z}\right),
\ee
\be
\Pi_0^{0i}(\bq,\o)\simeq\frac{n}{\o}\frac{q_i}{m_i},
\ee
\be
\Pi_0^{ij}(\bq,\o)\simeq\frac{n}{m_i}\d_{ij}.
\ee
The first expansion is exactly Eq.\ \pref{piooo} of the main text. If we substitute the last two into the definitions of $\Pi_T^{0J}$ and $\Pi_T^{JJ}$ (i.e. Eq.\ \pref{PiTOJJJ}) we get exactly Eqs.\ \pref{piojt} and \pref{pijjt}.

\section{Derivation of Eq.\ \pref{Pioo}}\lb{appc}

Let us consider once again the full density-density response function, i.e. Eq.\ \pref{beyondrpadd} for time-like indices $\mu=\nu=0$, that reads
\be
\lb{Pioofull}
\tilde{\Pi}^{00}(q)=\Pi_{RPA}^{00}(q)-\Pi_{MIX}^{0i}(q)
\L_{ij}^T(q)\Pi_{MIX}^{0j}(q).
\ee
The non-zero projection of $\Pi_{MIX}^{0i}$ along the direction set by $\hv_T^{xz}$, i.e.
\bea
(\hv_T^{xz})_i\Pi_{MIX}^{0i}(q)=
(\hv_T^{xz})_i\Pi_{RPA}^{0i}(q)=
\frac{\Pi_T^{0J}(q)}{1-V_C(\bq)\Pi_0^{00}(q)}\nn\\
\eea
has the crucial consequence that the mixing contribution $\Pi_{MIX}^{0i}\L_{ij}^T\Pi_{MIX}^{0j}$ in Eq.\ \pref{Pioofull} is in general non-zero. Indeed
\begin{widetext}
\bea
\lb{mixterm}
\Pi_{MIX}^{0i}(q)\L_{ij}^T(q)\Pi_{MIX}^{0j}(q)&=&
\left((\hv_T^{xz})_i\Pi_{MIX}^{0i}(q)\right)^2
\L_{xz}^T(q)\nn\\
&=&\frac{1}{1-V_C(\bq)\Pi_0^{00}(q)}
\frac{\left(\Pi_T^{0J}(q)\right)^2}
{\left(1-V_C(\bq)\Pi_0^{00}(q)\right)
\left(\Pi_T^{JJ}(q)-D_T^{-1}(q)\right)+
V_C(\bq)
\left(\Pi_T^{0J}(q)\right)^2},\nn\\
\eea
where we took into account the fact that the projections of $\Pi_{MIX}^{0i}$ along the longitudinal and the transverse $y$ direction are zero, as follows from $(\hv_L)_i\Pi_{MIX}^{0i}=(\hv_T^y)_i\Pi_{MIX}^{0i}=0$, while the one along the transverse $xz$ direction is finite. In Eq.\ \pref{mixterm} $\L_{xz}^T\equiv(\hv_T^{xz})_i(\hv_T^{xz})_j\L^{ij}=
1/\left(\Pi_{RPA}^{T,xz}-D_T^{-1}\right)$ is the transverse $xz$ component of $\L$, with $\Pi_{RPA}^{T,xz}
\equiv(\hv_T^{xy})_i(\hv_T^{xy})_j\Pi_{RPA}^{ij}=\Pi_T^{JJ}+\left(\Pi_T^{0J}\right)^2/\left(1-V_C\Pi_0^{00}\right)$. $\Pi_0^{0J}$ and $\Pi_0^{JJ}$ are the bare functions defined in Eq.\ \pref{PiTOJJJ} of the main text, which, along with the bare density function $\Pi_0^{00}$, allow for the following expression of the full density-density function:
\be
\lb{Tildepioo}
\tilde{\Pi}^{00}(q)=
\frac{1}{V_C(\bq)}
\left[
\frac{D_T^{-1}(q)
-\Pi_T^{JJ}(q)
}
{\left(
1-V_C(\bq)\Pi_0^{00}(q)
\right)
\left(D_T^{-1}(q)
-\Pi_T^{JJ}(q)
\right)-
V_C(\bq)
\left(\Pi_T^{0J}(q)\right)^2}-1
\right].
\ee 
\end{widetext}
Eq.\ \pref{Tildepioo}, that is valid at arbitrary momentum, is expressed in terms of the bare electronic susceptibilities $\Pi_0^{00}$, $\Pi_0^{0J}$ and $\Pi_0^{JJ}$ and can be easily recast in terms of the retarded density function $\Pi_{ret}^{00}$ defined in Eq.\ \pref{tildePioo}, see Eq.\ \pref{Pioo} of the main text. 

A last comment is in order about the undamped nature of the two modes $\o_-$ and $\o_+$ at $|\bq|\sim q_c$. To discuss it we plot the region identified by the values of $\bq$ and $\o$ for which ${\Pi_{ret}^{00}}''\neq 0$, where particle-hole (p-h) damping is operative, at three fixed values of the angle $\eta$ in Fig.\ \ref{fig3} ; we also show, in each panel, the corresponding frequency dispersions $\o_-(|\bq|,\eta)$ and $\o_+(|\bq|,\eta)$ as functions of $|\bq|$. Clearly both modes do not enter the continuum at $|\bq|\le q_c\simeq 5$ $\m$m$^{-1}$, i.e. they propagate with zero damping within the momentum range where mixing effects are relevant. Above the crossover value $\o_+\simeq\o_T\sim \tc|\bq|$ disperses as a pure transverse mode and never undergoes dissipation; on the other hand $\o_-(|\bq|,\eta)$ first saturates to its RPA value $\o_L(|\bq|,\eta)$ and then falls into the continuum.

\bibliography{Literature.bib}

\end{document}